\documentclass[nolinenumbers,trackchange]{aastex701}

\begin{document}

\title{Origin of Hyperion and Saturn's Rings in A Two-Stage Saturnian System Instability}

\author{Matija \'Cuk}
\affiliation{SETI Institute}
\email[show]{mcuk@seti.org}  

\author{Maryame El Moutamid}
\affiliation{Southwest Research Institute}
\email{maryame.elmoutamid@swri.edu}  

\author{Jim Fuller}
\affiliation{California Institute of Technology}
\email{jfuller@caltech.edu}

\author{Val\'ery Lainey}
\affiliation{LTE, Observatoire de Paris}
\email{valery.lainey@obspm.fr}

\begin{abstract}
The age of the rings and some of the moons of Saturn is an open question, and multiple lines of evidence point to a recent (few hundred Myr ago) cataclysm involving disruption of past moons. The main driver of the evolution of the Saturnian system is the relatively rapid tidal expansion of its largest moon, Titan, which is likely driven by resonant tides within Saturn. The obliquity of Saturn and the orbit of the small moon Hyperion both serve as a record of the past orbital evolution of Titan. Saturn's obliquity was likely generated by a secular spin-orbit resonance with the planets, while Hyperion is caught in a mean-motion resonance with Titan, with both phenomena driven by Titan's orbital expansion. We propose that the breaking of Saturn's spin resonance was also the event in which Hyperion formed, when an outer mid-sized satellite (``Proto-Hyperion'') was destabilized and collided with Titan, with some of the debris accreting into Hyperion. During the instability Proto-Hyperion's perturbations produced the observed orbital inclination of Iapetus. The same event also excited the eccentricity of Titan, which then, through Titan's resonant interaction with the inner moons, led to destabilization, collisional disruption and re-accretion of the inner moon system, including the rings. We present numerical integrations that show that this chain of events has a relatively high probability, and discuss how it fits within our knowledge of the Saturnian system.
\end{abstract}



\section{Introduction}\label{intro}

Saturn has a uniquely massive and bright ring system compared to those of the other three gas giants in the Solar System. A relatively young age of approximately 100 Myr for Saturn's rings has long been proposed based independently on dynamical timescales for ring-moon interaction \citep{gol82, lis85b} and compositional constraints \citep{cuz98}. Rapid orbital evolution of the moons close to the rings due to ring-moon gravitational interactions may indicate that the current architecture of the rings \citep[including the Cassini Division,][]{bai19} is young, but is less constraining about the overall age. Compositional arguments are based on the constant influx of dark cometary dust onto the rings and time needed for that dust to darken the otherwise icy rings. A crucial parameter determining the darkening timescales is the total mass of the rings, which determines the total amount of dust needed to pollute it \citep{cuz98}. A low ring mass was finally directly detected by Cassini Grand Finale data \citep{ies19}. Combined with the measurements of interplanetary dust flux \citep{kem23} these results imply an age of few hundred million years at most \citep{est23, dur23}, but there is still doubt among some researchers \citep{cri19, hyo25}. A similarly young age ($<100$~Myr) was also argued for Saturn's inner moons out to Rhea \citep{cuk16}, with a single dynamical instability being proposed as the origin of the inner moons and the rings. \citet{cuk16} argue that inner moons' orbits have too little eccentricity and inclination, as their expected orbital evolution over the age of the Solar System \citep{lai12} would result in accumulation of resonant excitations. 

More recently, it has been shown that orbital evolution of Titan and Rhea is more rapid than previously thought \citep{lai17, lai20, jac22}, suggesting that tidal evolution of these moons is driven by resonant response by the planet \citep{ful16} rather than equilibrium tides. If every single moon were in resonance lock with a Saturn's internal mode, the moons' would be on divergent or parallelly evolving orbits \citep{ful16, lai20}. However, the existence of resonances in the inner system (among the Mimas-Tethys and Enceladus-Dione pairs) requires convergent orbital evolution and therefore indicates a that not every single moon can currently be in resonance lock with internal modes. Subsequently, direct numerical modeling of recent resonances between the moons \citep{cuk23} and the discovery of an ocean within otherwise inactive Mimas \citep{lai24} both suggested the evolution timescale of tens of Myr for the dynamical evolution of inner moons. While it is impossible to definitely constrain the age of the inner moons based on any one mean-motion resonance (MMR) crossing due to the uncertain dependence of tidal response of Saturn on frequency, ages of billions of years appear unlikely at this point. Additional constraints can be inferred from interactions between Rhea and the solar evection resonance, which we will discuss in Section \ref{rhea}. Collision modeling \citep{teo23} has shown that a collisional disruption of the inner moons similar to Dione and Rhea (with the moons subsequently re-accreting) could also provide enough material to form the rings, as originally proposed by \citet{cuk16}.

Rapid tidal evolution of Titan has major implications for the outer Saturnian system. Obliquity of Saturn was almost certainly increased over time due to spin-orbit resonance between Saturn's axial precession and one of the nodal frequencies of the Solar System \citep{war04, ham04}. While originally thought to be a result of a slow and smooth planetary migration, currently the only plausible cause of the evolution into this resonance is the migration of Titan \citep{sai21a,sai21b}. Prior studies generally assumed that the spin-orbit resonance is still active, in part due to approximate alignment between the longitude of Saturn's equinox and the resonance's stable libration island \citep{war04, ham04, sai21b}. However, recent modeling of Saturn's interior implies that Saturn's spin is close to, but not in, the spin-orbit resonance with the solar system secular mode  $s_8$ \citep{wis22, man23}. This lack of current resonance was the basis of a model that proposes that the loss of an additional satellite, located between Titan and Iapetus, kicked Saturn out of the spin-orbit resonance about 100-200~Myr ago, with the additional moon (``Chrysalis'') being tidally disrupted during a close approach to the planet, giving rise to Saturn's rings \citep{wis22}.

In this paper we will present a dynamical model of the breaking of Saturn's spin orbit resonance that is similar to that of \citet{wis22} in involving a loss of an additional satellite, but with important differences. We will show  that the constraints from the current Titan-Hyperion orbital resonance imply a recent origin of Hyperion (Section \ref{hyp}), and we will argue that Hyperion must have originated in the same instability that broke the Saturn's spin-orbit resonance, requiring the additional moon to collide with Titan (Section \ref{outer}). We will also show that Rhea is not primordial satellite (Section \ref{rhea}), and that the inner moons are easily destabilized by an eccentric Titan following the initial instability, enabling the inner moons' re-accretion and formation of the rings \citep[cf.][]{teo23}. Finally, we will show that this scenario can produce otherwise unexplained inclination of Iapetus (Section \ref{iapetus}).

\section{The Orbit and Age of Hyperion}\label{hyp}

Titan and Hyperion are currently in the 4:3 mean-motion resonance, which keeps their orbital periods fixed in a 3:4 ratio. As Hyperion's orbit is relatively eccentric ($e=0.104$), the resonance is important for protecting Hyperion from having close encounters and colliding with Titan. 

It is established that the eccentricity of Hyperion in the resonance grows as Titan migrates outward \citep{md99}. In the context of equilibrium tides, the orbital evolution of Hyperion on its own is negligible, and this is even more true if Titan is resonance-locked. Therefore, the eccentricity of Hyperion could be used to constrain the past orbital evolution of Titan. According to Eq. 8.242 in \citet{md99}, the increase in the eccentricity of Hyperion is given by:
\begin{equation} 
{\dot{e}_H \over e_H} = {1 \over e_H^2} {m_T \over m_S} n_H a_H {F \over 3 g}
\label{md1}
\end{equation}
where $m$,$a$, $e$ and $n$ are masses, semimajor axes, eccentricities and mean motions, and subscripts $S$, $T$, and $H$ refer to Saturn, Titan and Hyperion. The other variables are defined as $F = 4 \dot{n}_H - 3 \dot{n}_T \approx -3 \dot{n_T}$, and $g=16 G m_T/a_H^2+9 G m_H/a_T^2 \approx 16 G m_T/a_H^2$, where $G$ is gravitational constant. Expanding $F$ and $g$ into Eq. \ref{md1}, and using $G m_S/a_H^3 = n_H^2$, $n_H=(3/4)n_T$ and $\dot{n}/n=-(2/3)\dot{a}/a$, we get:
\begin{equation}
2 e_H \dot{e}_H = {{\rm d} (e^2) \over {\rm d}t} = {1 \over 4} {\dot{a}_T \over a_T}
\label{md2}
\end{equation}
Assuming that Hyperion started with a low eccentricity, and evolved to $e_H=0.1$, Titan's semimajor axis must have evolved by about 4\%, assuming a constant rate of migration. This is a very rough estimate, but it gives us the correct order of magnitude of how much Titan could have migrated since Hyperion was captured into the resonance.


When \citet{lai12} presented evidence for the fast tidal evolution of Saturn's moons, \citet{cuk13} noted that the new value of Saturn's tidal quality factor $Q \approx 1700$ (with Saturn's tidal Love number $k_2=0.37$) would make the Titan-Hyperion resonance about as old as the Solar System \citep[cf.][]{gre73}. The more recent direct measurement of Titan's orbital evolution \citep{lai20} puts the effective tidal $Q$ of Saturn at Titan's frequency as $Q \approx 120$, equivalent to an evolution timescale of $a/\dot{a}=11$~Gyr. Such a rapid expansion of Titan's orbit would place the origin of the Titan-Hyperion resonance at about 400-500~Myr ago.

\begin{figure}[ht]
\plotone{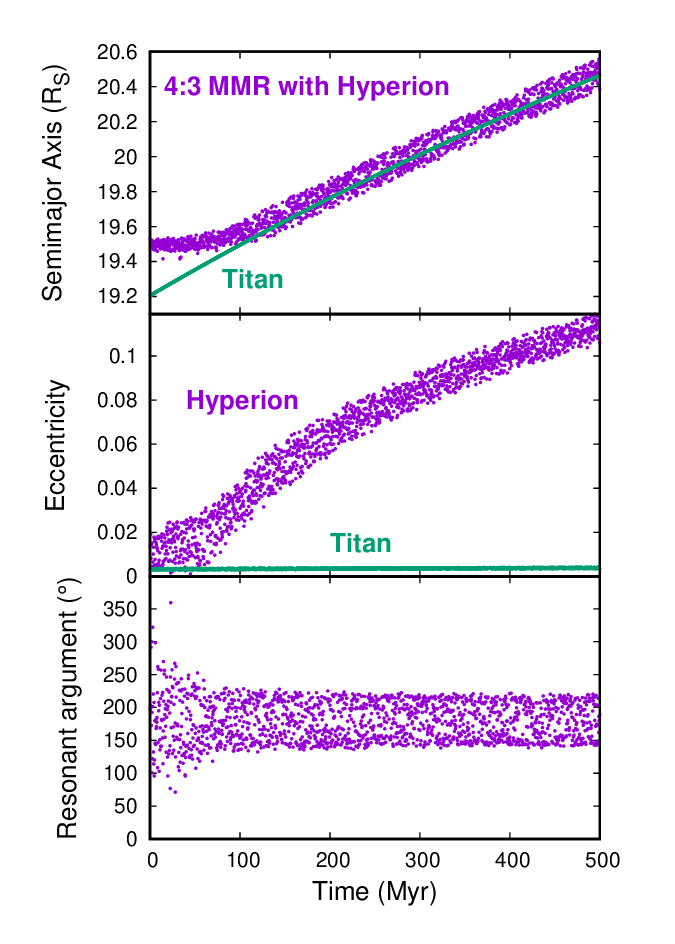}
\caption{Simulations of Titan and Hyperion encountering and evolving through their mutual 4:3 mean-motion resonance, replotted from \citet{cuk13}. The top panel plots Titan's semimajor axis (green) and the location of the 4:3 MMR with Hyperion (purple; calculated as $a_H (3/4)^{2/3}$), and the middle panel plots their eccentricities. The resonant argument plotted in the bottom panel is $4 \lambda_H - 3 \lambda_T - \varpi_H$, where $\lambda$ is the mean longitude and $\varpi$ the longitude of pericenter. The current proper eccentricity of Hyperion (comparable to the one plotted in this figure) is $e_H=0.1$. Titan was put on a low-eccentricity orbit for more direct comparison with analytical calculations.\label{hyp_plot}}
\end{figure}

Numerical integrations support the above calculations. Figure \ref{hyp_plot} shows that Hyperion's eccentricity requires an orbital evolution of Titan of about 5\%. This simulation used the standard version of the mixed-variable symplectic integrator {\sc simpl} \citep{cuk16} with artificially accelerated equilibrium tides, but given that the main dynamical mechanism in this resonance is the relatively uniform expansion of Titan's orbit, it can be directly compared with the MMR evolving under resonance lock. The rate of Titan's orbital expansion in the simulation corresponds to tidal quality factor of $Q=100$ for Saturn, putting this run on the higher end of the expansion rate found by \citet{lai20}, confirming the model age of the resonance to be 400-500~Myr. The actual age of Hyperion itself must be somewhat greater, but our simulations show that Hyperion had to form interior to 7:5 MMR with Titan, otherwise it would have been captured into that resonance. Furthermore, preliminary runs show that Titan’s 11:8 MMR gives Hyperion sizable kicks in eccentricity ($\approx 0.04$) inconsistent with Hyperion’s current orbit, indicating that Hyperion had to form interior to 11:8 MMR, too. Therefore, pre-resonance lifetime of Hyperion had to be less than 200 Myr, but could also have been a lot less.   


Here we assume negligible tidal migration by Hyperion itself, as for Hyperion to be in resonance lock, Saturn must have $k_2/Q=250$ at Hyperion's orbital frequency, which is nonphysical.  The question of eccentricity damping within Hyperion is complicated by Hyperion's chaotic rotation \citep{wis84}. For bodies in synchronous rotation, damping of eccentricity is given by \citet{md99}:
\begin{equation}
\dot{e} = - e {21 k_2 \over 2 Q} {m_{planet} \over m_{moon}} \left({R_{moon} \over a}\right)^5 n
\label{edot}
\end{equation}
If we use the most optimistic tidal Love number $k_2=0.01$ for Hyperion (a demonstrably solid, irregularly shaped object) and tidal quality factor $Q=10$, as well as $R_H=135$~km, we get the timescale $e/\dot{e}=10^{12}$~yr for eccentricity damping. This lower limit was likely valid when the eccentricity of Hyperion was low early in the resonance, as chaotic rotation is driven by orbital eccentricity \citep{wis84}.

In chaotic rotation Hyperion does not follow regular librations due to eccentricity, but has a practically randomized orientation relative to Saturn. Now energy dissipation is better approximated as either a tidal spindown, or damping of non-principal axis rotation \citep[as Hyperion's rotation is chaotic in all three dimensions,][]{bla95}. The rate of spindown is given by \citep{md99}:
\begin{equation}
\dot{\omega} = {3 k_2 \over 2 C Q} {m_{planet} \over m_{moon}} \left({R_{moon} \over a}\right)^3 n^2
\label{omegadot}
\end{equation}
where $C$ is the moment of inertia (we will assume $C=0.5$ for highly-elongated Hyperion). Using the same very optimistic assumptions as before, we get $\dot{\omega}/n \approx 10^5$~yr, meaning that the energy of chaotic rotation that is roughly $n$ away from synchronous rotation rate is damped on that timescale. However, to damp the eccentricity, we need to decrease the semimajor axis of Hyperion by $a e^2$, which means changing its orbital energy by a fraction $e^2 \approx 0.01$. Orbital energy is larger than rotational by about the factor of $(a/R_H)^2 \approx 10^8$. Therefore even in chaotic rotation the eccentricity damping timescale should be at least $10^{11}$~yr.

Alternatively, if we assume internal dissipation driven by non-principal axis rotation damping, we can use the estimate of the timescale of this process calculated by \citet{sha05}:
\begin{equation}T_{NPA}= 0.24 \ {\rm Myr} {(P/1 \ {\rm hr})^3 \over (R/1 \ {\rm km})^2} \left({ \mu Q \over 5 \times 10^{11} {\rm N \  m^{-2}}}\right) \left({2500 \ {\rm kg \ m^{-3}} \over \rho }\right)
\label{sharma}
\end{equation}
where $P$, $R$, $\mu$ and $\rho$ are the body's rotation period, radius, rigidity and density, respectively. Our optimistic choice of $Q/k_2=10^3$ is equivalent to $\mu Q \approx 10^9$~N~m$^{-2}$. Using $\rho_H=500$~kg~m$^{-3}$ and assuming the rotation frequency that is twice the synchronous rate, we get $T_{NPA} \approx 2$~Myr, putting the eccentricity damping timescale on the order of $10^{12}$~yr. We conclude that Hyperion's eccentricity damping is negligible on sub-Gyr timescales (see Appendix \ref{app} for additional discussion).


Hyperion is a relatively small moon, but its relatively recent origin would very likely imply a wider cataclysm at the time of its formation. \citet{ham13} has proposed that Titan was a late merger between multiple satellites, with Hyperion being an unaccreted fragment. \citet{asp13} also proposed a late formation mechanism for the Saturnian moons that involved a major impact on Titan, which they similarly propose to be a late merger. The strong indications of Hyperion’s relatively young age support the hypothesis that it formed during a late, system-wide cataclysm involving Titan, potentially through a major impact or merger event, rather than being a primordial remnant.

\section{Outer System Instability}\label{outer}

The Chrysalis hypothesis \citep{wis22} helps explain not only the breaking of the spin-orbit resonance, but also naturally produces the high inclination and eccentricity of Titan (excited by Chrysalis  through their mutual resonance) which would otherwise be damped by Titan's tides if it were older than about 350~Myrs \citep{dow25}. Simultaneous survival of Hyperion-Titan resonance and the tidal disruption of Chrysalis is an unlikely, though still plausible, outcome of the initial conditions proposed by \citet{wis22}. Our own simulations aimed at reproducing the numerical experiments of \citet{wis22} using a different numerical approach \citep[a simple T+U symplectic integrator {\sc complex};][]{for90, cuk16} agree with their results. We do find that the survival of Hyperion, either in and outside the resonance with Titan, is sensitively dependent on the eccentricity of Hyperion at the time \citep[][ assumed initial $e_H=0.045$]{wis22}\footnote{The requirement for a much lower eccentricity of Hyperion few hundred Myr ago makes the idea of a long-term constant eccentricity of Hyperion \citep{gol24a} difficult to reconcile with the Chrysalis hypothesis of \citet{wis22}.}. However, as we have shown in the previous section, the mean-motion resonance between Titan and Hyperion appears to be only about 400 Myr old, during which time the eccentricity of Hyperion increased approximately as $e_H \propto t^{1/2}$. Therefore at the time of the Chrysalis instability Hyperion itself could not have been more than a couple of hundred Myr old, implying either existence of a previous, separate event, or that the hypothesis may need modification. Here we propose that Hyperion did not survive the loss of an additional moon exterior to Titan, but that Hyperion formed in this event. We show that this version of the instability, which then spreads to the inner system, can better explain the peculiarities of the Saturnian system through higher-probability events than in the previous proposals.

\begin{figure}
\plotone{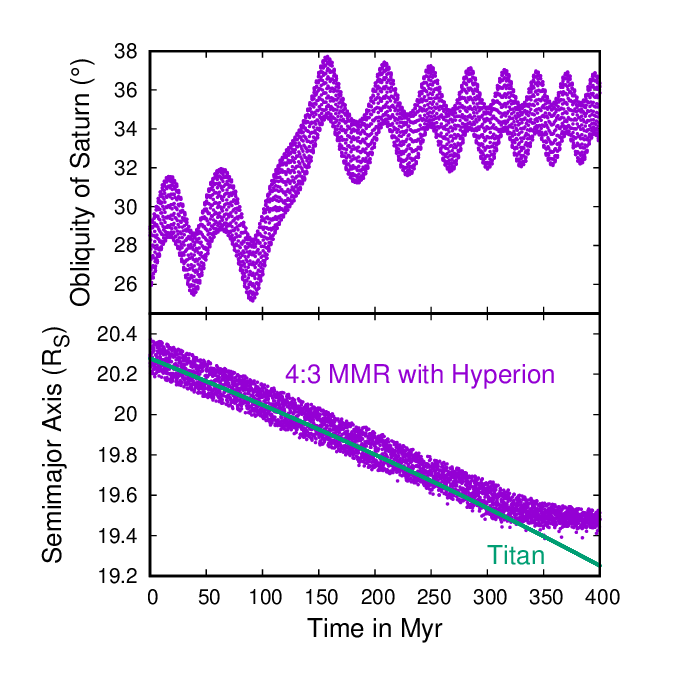}
\caption{The numerical experiment using {\sc ssimpl} in which Titan was made to migrate inward at the same rate it is found to migrate outward by \citet{lai20}. The top panel plots the obliquity of Saturn, while the bottom one shows Titan's semimajor axis (green) and the location of the 4:3 MMR with Hyperion (purple; calculated as $a_H (3/4)^{2/3}$). While not a reverse-time simulation, this run shows approximate past points in Titan's evolution when the spin-orbit resonance was crossed (indicated by a large jump in Saturn's obliquity) and Titan-Hyperion resonance was established (consistent with the forward-time simulation shown in Fig. \ref{hyp_plot}).\label{reverse}}
\end{figure}

Figure \ref{reverse} illustrates how the obliquity of Saturn and the orbit of Hyperion are dynamically coupled to the tidal evolution of Titan over the past few hundred Myr \citep[here we used mixed-variable symplectic integrator {\sc ssimpl} which evolves the planetary spin-axis precession while assuming a constant spin rate,][see also Appendix \ref{app_b}]{cuk18}. Assuming Titan's observed migration rate \citep{lai20}, Saturn should have crossed the spin obit resonance 100-200 Myr ago, while Hyperion entered its current orbital resonance with Titan only about 400 Myr ago. Breaking of the spin-orbit resonance through a loss of a Saturnian satellite would make Saturn's spin axis precess slower than required by the resonance \citep[right now spin precession is too fast;][]{wis22}, so it could not have happened less than 100-200 Myr ago. On the other hand, Hyperion could not have resided on a non-resonant orbit for long, as it would have been captured by mean-motion resonances with Titan other than the present 4:3 MMR. The disappearance of a major moon exterior to Titan and the appearance of a smaller, irregularly shaped and highly porous \citep{tho07} Hyperion in the same timeframe are likely to be related.


The most likely outcome for "Chrysalis" (about 50\%) in our simulations reproducing \citet{wis22} was not tidal disruption by Saturn but collision with Titan or Hyperion. Furthermore, collisions between Titan and Hyperion were very common in the runs in which we initially placed Hyperion deeper in the resonance with Titan (i.e. with $e_H>0.05$), as this eccentricity would put Hyperion on Titan-crossing orbit if it were to be ejected from resonance by encounters with Chrysalis. We find that this implies that the Chrysalis hypothesis is incomplete, as it does not explain the extremely recent resonant capture of Hyperion.

We also run simulations without Hyperion and found very similar outcomes, with ejection and collisions with Titan being the most common outcomes for Chrysalis, with collisions with Iapetus being less likely (at about 10\% level). While Iapetus was often strongly perturbed by Chrysalis, the orbital element of Iapetus that was {\it least} affected was typically Iapetus's inclination, compared to its semimajor axis and eccentricity.

\begin{figure}
\plotone{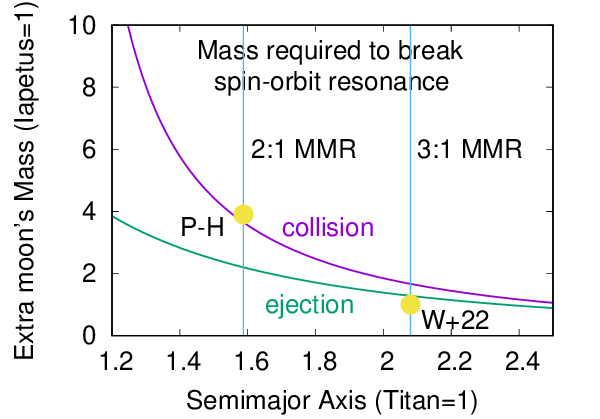}
\caption{Distance vs. mass of hypothetical additional satellite needed to put Saturn in the spin-orbit resonance with secular mode $s_8$ in the past.\label{resonances}}
\end{figure}

Therefore we formulated an alternative but closely related hypothesis of "Proto-Hyperion" which is also a past moon orbiting between Titan and Iapetus. As Proto-Hyperion gives rise to Hyperion, Hyperion is not initially present in our setup. In order to avoid close encounters with Iapetus in our simulations, we placed Proto-Hyperion just outside the outer 2:1 resonances with Titan, rather than the 3:1 MMR that was proposed by \citet{wis22} for Chrysalis. With that semimajor axis, increasingly eccentric Proto-Hyperion would have its pericenter encounter Titan's orbit well before its apocenter approaches that of Iapetus. A smaller semimajor axis of the additional moon means that it needed to have a larger mass in order to contribute sufficient effective $J_2$ moment to enable Saturn's spin orbit resonance in the past (Fig. \ref{resonances}). If the additional satellite is removed from the outer system (either becoming unbound or forming the rings), the mass of the additional satellite depends on its semimajor axis as $m \propto a^{-2}$ in order to produce the same effective $J_2$ (green line in Fig.\ref{resonances}). However, a satellite that merges with Titan will (in order to produce the same {\it additional} effective $J_2$) have dependence of mass on orbital distance as $m \propto (a^2-a_T^2)^{-1}$ (purple line in Fig. \ref{resonances}), as the extra moon's mass is subsequently added to Titan's. 
Therefore the mass required by Proto-Hyperion is about four times that of Chrysalis. We used $3.6 \times 10^{-9} M_{Sun} \approx 1.26 \times 10^{-5} M_{Saturn} \approx 7.2 \times 10^{21}$~kg as the assumed mass of proto-Hyperion, and we reduced the initial mass of Titan by the same amount. 

We initially placed Titan at $a=19 R_{S}$ and proto-Hyperion just outside Titan's outer 2:1 resonance, while Iapetus was placed at its current semimajor axis. All three moons were given small eccentricities $e \leq 0.001$ and free inclinations $i_{free} \leq 0.1^{\circ}$ (all three moons have significant forced inclinations). The obliquity of Saturn was set to be constant at $37^{\circ}$, as the integrators we used in this case ({\sc simpl} and {\sc complex}) did not evolve the planet's spin. While the tidal evolution of Titan is almost certainly driven by resonance-lock tides \citep{ful16, lai20} in our simulations we used an equilibrium tidal model with tidal quality factor $Q=100$ and Love number $k_2=0.37$ for Saturn, as it is much easier to implement than resonance-lock tides. This shortcut was justified as Titan is the only moon experiencing significant tidal evolution in this setup, and later in this Section we discuss whether Titan's resonance lock would survive the instability. We turned off satellite tides in {\sc simpl} as they were unlikely to produce any noticeable effects during the instability, especially if pre-instability Titan that was on a circular and planar orbit had a much more rigid interior. Gravitational perturbations from Sun on the moons and from Jupiter on Saturn were also included in the simulation.

The three-moon system described in the last paragraph was integrated using {\sc simpl} until orbit-crossing was detected about 54~Myr into the simulation. We then used the dynamical state of the system at 53.5~Myr to start an array of 60 simulations using a more basic symplectic integrator {\sc complex}, which splits the Hamiltonian into the kinetic and potential energy, rather than Keplerian motion and perturbation, as {\sc simpl} does. This lack of Keplerian assumption makes {\sc complex} slower but also better suited for situation with crossing orbits. The 60 simulations were all started with identical initial conditions but each had a slightly different timestep in the $2.7-3.3 \times 10^{-5}$~yr range (the first-stage simulation in {\sc simpl} had a $10^{-3}$~yr timestep), which was reduced by a factor of 20 when two moons were less than 10 combined radii from each other. The simulations were run until the moons collided with each other or the planet, or one of the moons was ejected from the system. 

\begin{figure}
\plotone{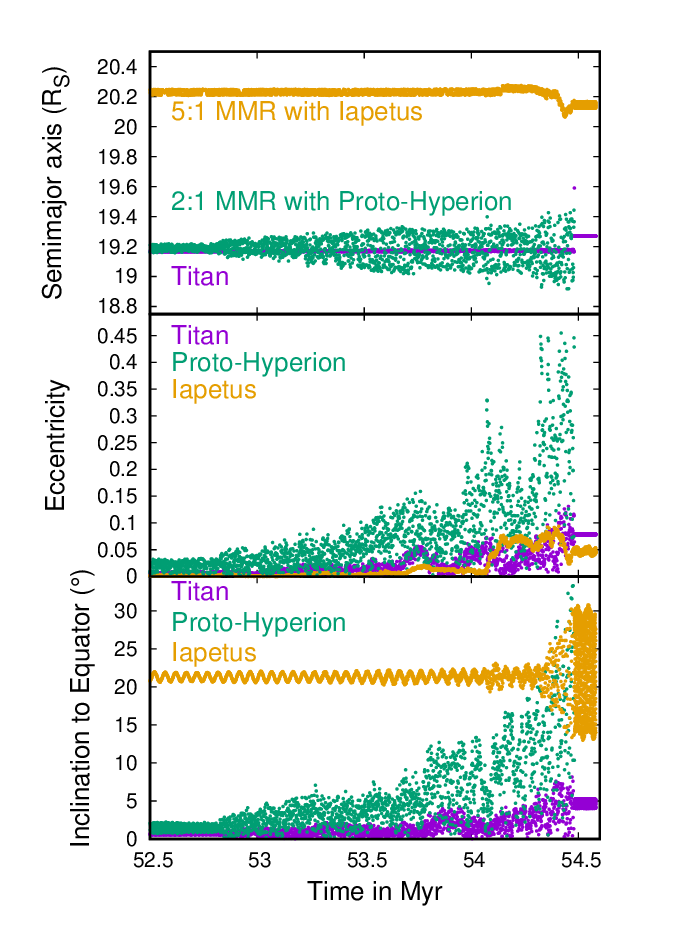}
\caption{A (successful) example of the evolution of the system in our simulations that place ``Proto-Hyperion'' in the outer 2:1 resonance with Titan. The top panel plots Titan's semimajor axis (purple) and the location of the 2:1 MMR with Proto-Hyperion (green; calculated as $a_{P-H} (1/2)^{2/3}$) and 5:1 with Iapetus (orange; calculated as $a_I (1/5)^{2/3}$). The middle and bottom panels plot the eccentricities and inclinations to Saturn's equator of the three moons using the same color scheme. \label{p-h4.4b57}}
\end{figure}

Out of 60 simulations that were all started just before Titan-Proto-Hyperion 2:1 MMR, proto-Hyperion collided with Titan in 42 simulations, while either proto-Hyperion or Iapetus were ejected in the remaining runs. In the simulations that ended up with Titan-Proto-Hyperion collision, Iapetus typically experienced extensive gravitational perturbations from the Proto-Hyperion during the chaotic phase of the latter's evolution. These perturbations often resulted in significant eccentricity and inclination of Iapetus's orbit that was initially close to being circular and in the Laplace plane. Figure \ref{p-h4.4b57} shows one simulation in which proto-Hyperion collided with Titan and Iapetus's final orbit resembles the current one. After the Titan-proto-Hyperion collision, we merged the two moons (conserving mass and momentum) and integrated the new system of full-sized Titan and Iapetus for $10^{5}$~yr to be able to calculate their average elements (all 42 systems were stable for this long).

\begin{figure}
\plotone{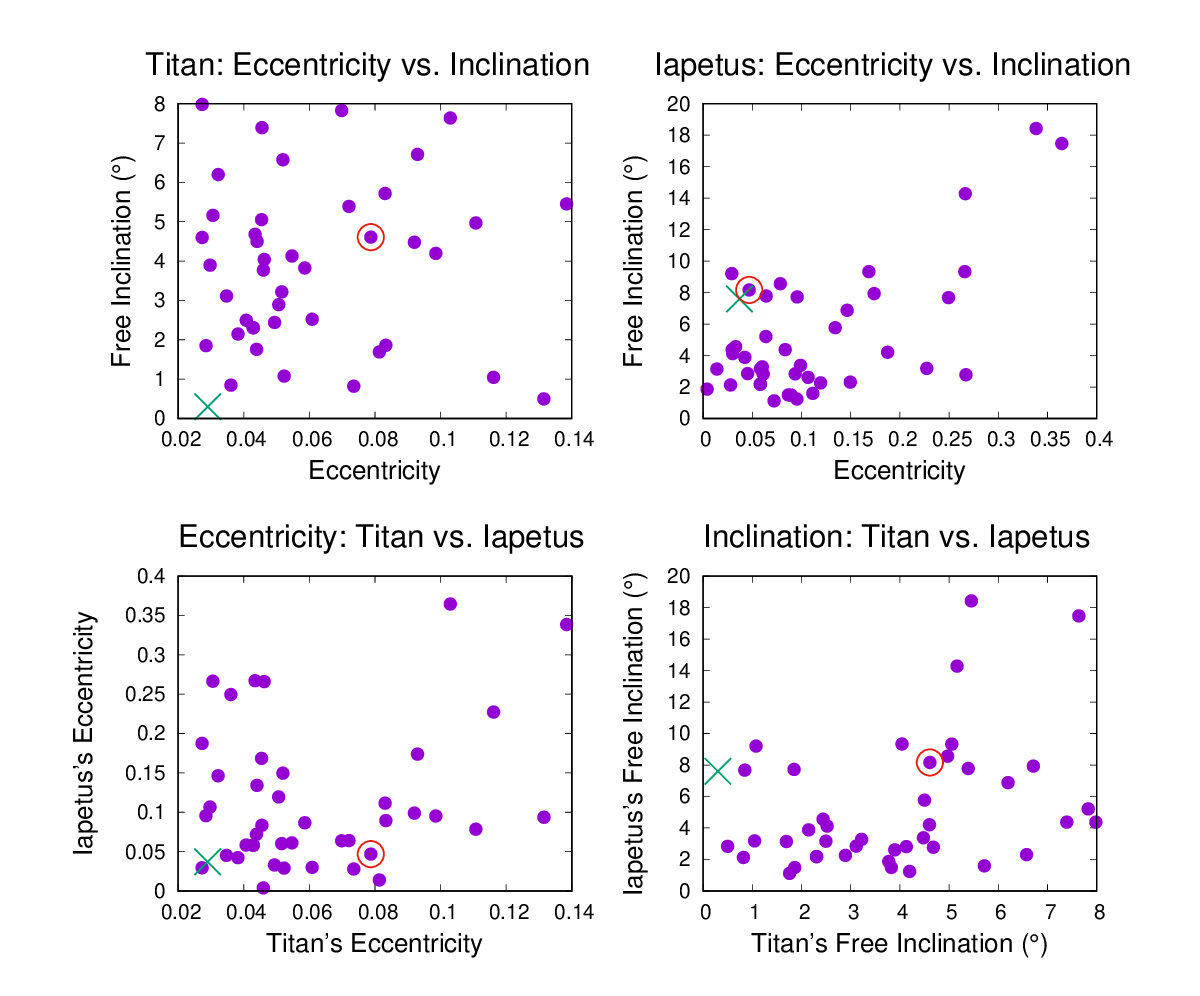}
\caption{Final eccentricities and inclinations of Titan and Iapetus in the 42 runs that resulted in a collision between Titan and Proto-Hyperion. The red circle indicates the simulation plotted in Fig. \ref{p-h4.4b57}, and the green "X" plots the current orbital elements. \label{all_p-h4.4}}
\end{figure}

Figure \ref{all_p-h4.4} shows the final eccentricities and free inclinations of Titan and Iapetus in the 42 ``successful'' simulations. The eccentricities and free inclinations of each moon are plotted against each other (top row) and against the same element for the other moon (bottom row). Apart from some of the most dynamically excited outcomes for Iapetus having both high eccentricity and inclination, there is no obvious correlation between any of the orbital elements. While the current orbit of Iapetus is within the wide range of possible outcomes, the current orbit of Titan is much less dynamically excited, which (qualitatively speaking) is a natural consequence of tidal dissipation \citep{dow25}. 

A more qualitative examination of tidal dissipation within Titan and consequences for our hypothesis suggests two distinct pathways to reconciliation. For the eccentricity of Titan $e_T=0.029$ to survive to the present day from an excitation 500~Myr ago, Titan's eccentricity damping timescale has to be about 500~Myr or more, which is an order of magnitude longer than determined by \citet{dow25}, or has recently changed so that the observed fast damping is not typical of Titan's post-excitation history. The first possibility would require \citet{dow25} modeling to be incomplete, and that most of the inclination damping indicated by Cassini state phase offset must come from friction between solid and fluid layers in the interior of Titan. While the source of the inclination damping does not change the inclination damping timescale, it does have consequences for the associated damping of eccentricity, which is faster than inclination damping for tide-driven damping and zero for ocean-shell boundary friction \citep{che14, dow25}. One argument for this interpretation is that the large offset of Titan's Cassini state obliquity from one expected for a solid body implies multiple decoupled components. 

The second possibility, that the tidal response of Titan has become much stronger in last few Myr is harder to constrain with present-day measurements, but may be necessary to reconcile both \citet{wis22} and our hypothesis with a current eccentricity damping timescale of 30~Myr for Titan. Recent results obtained by \citet{pet25} support this very fast tidal dissipation and also propose to explain the obliquity of Titan without requiring an ocean. If the eccentricity of Titan has a different source from an outer system instability discussed here \citep[or by ][]{wis22} for which it is a natural outcome, an even more recent source of excitation would be needed. Neither the inner system cataclysm (Section \ref{inner}) nor 5:1 MMR with Iapetus (Section \ref{iapetus}) appear to be able to directly increase Titan's eccentricity significantly, although \citet{cuk16} do propose an indirect connection to the inner-system re-accretion though interactions with the outer edge of the $\approx 10 R_S$ disk, a mechanism that has not been demonstrated to work yet using numerical simulations.


\begin{figure}
\plotone{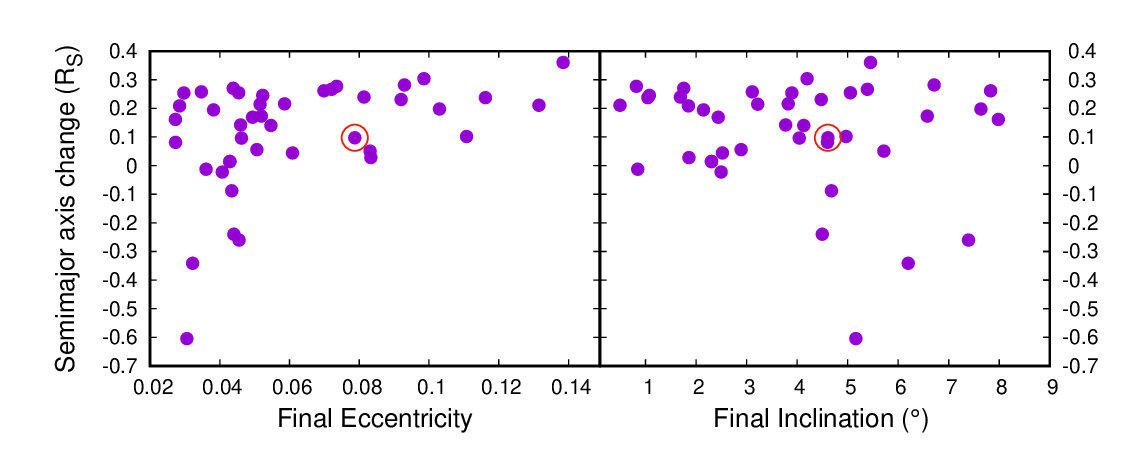}
\caption{The difference between the final semimajor axis of Titan and one expected from smooth tidal evolution plotted against final eccentricities and inclinations of Titan in the 42 runs that resulted in a collision between Titan and Proto-Hyperion. Red circle indicates the simulation plotted in Fig. \ref{p-h4.4b57}.}
\label{jump}
\end{figure}

One important question regarding this scenario is that of the continuity of Titan's resonance lock. During the merger Titan experiences an impulsive change in orbital elements, including the semimajor axis. If the semimajor axis increases, the resonance lock will likely reestablish itself after a period of slower orbital migration as the mode  catches up with Titan. However, a negative change in the semimajor axis may put Titan interior to the orbit corresponding to the resonance lock, permanently breaking the lock.  Figure \ref{jump} plots the size of Titan's semimajor axis change against its eccentricity and inclination in the Titan-Proto-Hyperion collision simulations. A large majority of these simulations result in an outward shift by Titan. This is understandable, as Titan is merging with Proto-Hyperion which is usually on an exterior orbit at the time of their collision. An outward jump of 0.1~$R_S$ would increase the reconstructed age of the cataclysm by about 50~Myr, or about 10\%. 

As we model the collision as a simple merger, we have no way of estimating how much ejecta would be produced and whether that ejecta can accrete into Hyperion. We did review close encounters between Proto-Hyperion and Titan before the final event (collision or ejection), and in a significant minority of simulations Proto-Hyperion may experience tidal disruption by Titan before the final collision. A more detailed exploration of these issues will require significant further collisional modeling. We do note that the crater population on Iapetus seems to be extremely ancient and dominated by heliocentric impactors, as Iapetus's crater size-frequency distribution \citep{rob24} matches that of craters on Charon \citep{sin19, rob21}. This fact may put an  upper limit on the amount of debris that was released during the demise of Proto-Hyperion and subsequently impacted Iapetus, but more work is needed to quantify this limit. The surface of Titan is estimated to be 0.2-1~Gyr old \citep{nei12}, which is consistent with resurfacing during the outer system instability proposed here. 

 While it is certain that some debris would be produced in Titan-Proto-Hyperion collision, it is not clear how the debris would decouple from Titan in order to form Hyperion. If Hyperion's material is mostly derived from Proto-Hyperion rather than Titan, this could be due to significant fraction of proto-Hyperion ``missing'' Titan in a glancing collision, but continuing to orbit on a modified orbit with semimajor axis larger than Titan's \citep[cf. ][]{can01, asp06}. If a minor fraction of material (probably consisting of small fragments) can collide into a ring before being re-accreted by Titan, that ring could accrete into a new moon. As the mass of Hyperion is $<0.1\%$ of Proto-Hyperion's mass, Hyperion's formation certainly did not have to be an efficient process. As the density of Hyperion ($\rho_H=540$~kg~m$^{-3}$) implies $>40\%$ porosity \citep{tho07}, Hyperion is very likely to be a ``rubble-pile'' than  a single solid fragment of a compacted icy moon. As for Titan, the merger with proto-Hyperion would clearly be a major turning point in the moon's evolution, with major implications for its surface and atmosphere \citep{nei12, man12}.

Finally, while the outcomes shown in Fig. \ref{all_p-h4.4} covers the phase space that includes the current orbit of Iapetus, direct comparison is complicated by subsequent evolution. Much more recently than the cataclysm discussed here (which happened 400-500 Myr ago), Titan and Iapetus crossed their mutual 5:1 mean-motion resonance \citep{pol18}. In Section \ref{iapetus} we address how the orbit of Iapetus evolved further due to this recent resonance with Titan. 

\section{Dynamics of Rhea and the Age of Inner Moons}\label{rhea}

\citet{cuk16}, assuming equilibrium-type tides and a tidal $Q \simeq 1700$ as found by \citet{lai12}, analyzed the orbital histories of the three largest moons interior to Titan: Tethys, Dione and Rhea. \citet{cuk16} found that their orbits are consistent with Dione and Rhea crossing their mutual 5:3 mean-motion resonance in the past, but that the orbital inclinations of the moons exclude their passage through the Tethys-Dione 3:2 MMR. Assuming $Q=1700$, this relative dynamical age translates to a young absolute age of the system of about 100 Myr. \citet{cuk16} concluded that the rings and moons interior to Titan formed in a dynamical instability about 100 Myr ago, in which the previous generation of moons was disrupted in collisions, and then largely reaccreted into the observed satellites.    

Non-dynamical arguments have also been used to constrain the inner moons' age. It appears that dominant past impactors in the Saturnian system are different from Kuiper Belt objects \citep{zah03, sin19}, and are possibly planetocentric \citep{fer20, bel20, fer22a, fer22b}. This would be consistent with, but would not require, a recent origin of the system. Recent confirmation of the existence of an ocean within Mimas \citep{taj17, lai24} implies a young ocean (in the tens of Myr) and still-evolving Mimas \citep{rho22, rho23, rho24a, lai24}. A young ocean within Mimas probably implies a recent origin of its orbital eccentricity, possibly in a resonance with another moon \citep{cuk22}. None of these findings require the inner moons of Saturn to be younger than some set age, but the fact that much of their evolution may have happened within the last 100~Myr would be consistent with a young age.

Since the publication of \citet{cuk16}, analysis of the moons' motions using both Earth-based astrometry and Cassini data \citep{lai17, lai20, jac22} has strongly suggested that the tidal evolution in the Saturnian system is not driven by equilibrium tides. Both Rhea and Titan have been found to migrate outward many times faster than predicted by equilibrium tides, with orbital evolution timescales of 6 and 11~Gyr, respectively. These tidal evolution timescales are much closer together than equilibrium tides would predict, and close to the tidal evolution timescales found for the closer-in moons (although with larger uncertainties). \citet{cuk23} have found that the observed orbital resonances (Mimas-Tethys 4:2 and Enceladus-Dione 2:1 MMRs) in the inner Saturnian system can be explained by a combination of equilibrium and resonant tidal evolution over the last $\approx 30$~Myr, reinforcing the idea that the inner moons may be young. 

The apparent rapid tidal migration of Rhea \citep[and apparent slow evolution of Tethys;][]{cuk23} help avoid the constraint from \citet{cuk16} that the inner moons must be younger than 100 Myr in order to avoid a past Tethys-Dione 3:2 resonance. However, the rapid evolution of Rhea's orbit offers us another way to constrain the age of Rhea. Assuming that Rhea's tidal evolution has been driven by resonance lock over most of the system's history \citep[this has to be true for most moons if they are to be several Gyr old][]{lai20}, Rhea should have crossed the evection resonance with the Sun within the last Gyr. 

Evection resonance is a semi-secular resonance in which a moon's apsidal precession period is equal to the parent planet's orbital period \citep{tou98}. Evection is considered a semi-secular resonance because the (apparent) mean motion of the perturber (the Sun) is in resonance with the apsidal precession of the satellite's orbit. As the moons' orbital precession is overwhelmingly driven by Saturn's oblateness, the dynamics of the evection resonance does not depend on the relationship between orbital periods of different moons. The evection resonance affects any moon at a certain distance from Saturn, which is given by:
\begin{equation}
a_{eve} = R \left({3 \over 2} J_2 {\Omega_0 \over n_s} \right) ^{2/7} 
\label{eve}
\end{equation}
Where $R$ is Saturn's radius, $J_2$ is Saturn's oblateness moment, $\Omega_0 =\sqrt{G M / R^3}$ is the orbital frequency at $a=R$, and $n_S$ is Saturn's heliocentric orbital mean motion. Eq. \ref{eve} gives as $a_{eve}=8.1 R$, but Titan's perturbations shift the location of the evection resonance to $a_{eve}=8.2 R$. This distance is somewhat smaller than the current orbital distance of Rhea ($a_R=8.7 R$). 

\begin{figure}
\plotone{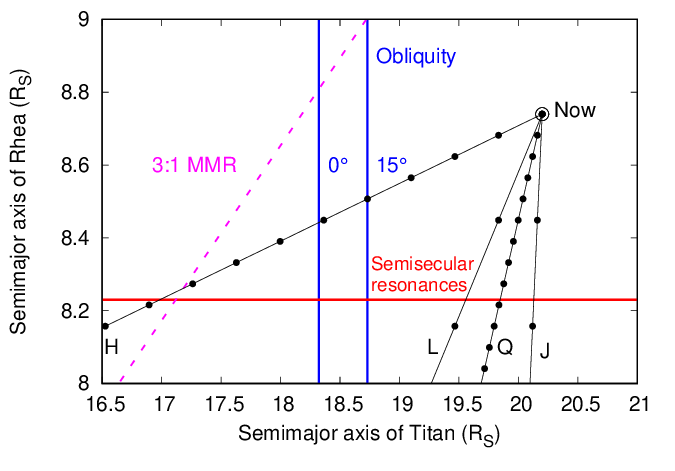}
\caption{Location of possible past dynamical features such as resonances plotted against past semimajor axes of Rhea and Titan. Black lines with dots plot possible past evolutionary tracks of the system, with each black dot indicating a 200~Myr step. The four lines are based on four possible scenarios: fast Rhea, fast Titan \citep[case L,][]{lai20}, fast Rhea, slow Titan \citep[case J, ][]{jac22}, relatively slow Rhea and Titan \citep[case Q,][]{lai12}, and slow Rhea, fast Titan (case H, purely hypothetical). \label{aa_plot}}
\end{figure}

Estimates of the past orbital evolution of Rhea through resonance locking \citep{lai20} suggest that Rhea should have crossed the evection resonance about 300-400 Myr ago. Of course, this timing depends on the question of whether Rhea was migrating at the same rate then as it does today. Figure \ref{aa_plot} plots the relevant phase space for various resonance involving Rhea and Titan in the plane of semimajor axes of Rhea and Titan. While here we concentrate on the migration rates found by \citet{lai20}(black line labelled ``L''), we also included the results found by \citet{jac22}(black line labelled ``J''). While these works differ in their measurements of Titan's migration rate, they both agree on the fast migration of Rhea and predict an evection resonance crossing by Rhea about 400~Myr ago. For completeness, we also included the path expected from equilibrium tides with $Q=1700$ \citep[][; black line labeled ``Q'']{lai12}, in which case Rhea crossed the evection resonance about 2 Gyr ago. 

The evection resonance, with the resonant argument $2 \lambda_S - 2\varpi$ ($\lambda_S$ is Saturn's mean longitude, $\varpi$ Rhea's longitude of pericenter), is expected to significantly excite Rhea's eccentricity. One possible outcome is that Rhea's satellite tides become strong enough to push the satellite through the resonant mode, leaving it stranded interior to the mode. Another possibility would be Rhea evolving with the mode, while captured in the evection resonance, producing a continuously increasing eccentricity. Yet another outcome would be a Rhea that continues evolving through the mode, but exits the resonance, with Rhea's excited  eccentricity slowly damping. Only the last of these possibilities would be consistent with the present orbital distance and observed evolution rate of Rhea.

In order to include resonance locking into {\sc simpl}, we need to include four new quantities. These are the initial frequency of the resonant mode $\nu$, its rate of change $\dot{\nu}/\nu$, peak dissipation of the resonant mode $1/Q_0$, and the width of the normal mode $\sigma$. The effective $Q$ for a satellite near a resonant mode is given by:
\begin{equation}
Q = Q_0 + \left( n - \nu \over \nu \sigma \right)^2
\label{nomode}
\end{equation}
where $n$ is the moon's mean motion. We used $Q_0=10$ and varied $\sigma$ in the $10^{-6} \leq \sigma \leq 10^{-5}$ range for the simulations shown in this Section. We chose $Q_0=10$ as a physically plausible value that is lower than any observed effective $Q$ (100-300 for Titan and Rhea), and this range of $\sigma$ was chosen as it gives us varied behavior at the evection resonance (resonance lock to even wider modes does not break, locking to even narrower modes is short-lived in our implementation). 

\begin{figure}
\plottwo{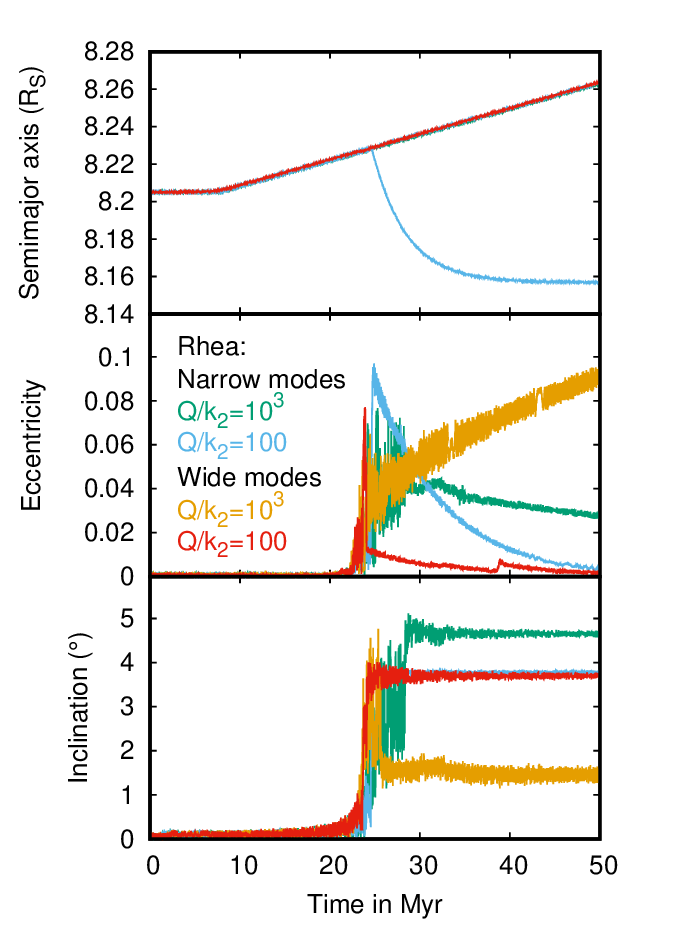}{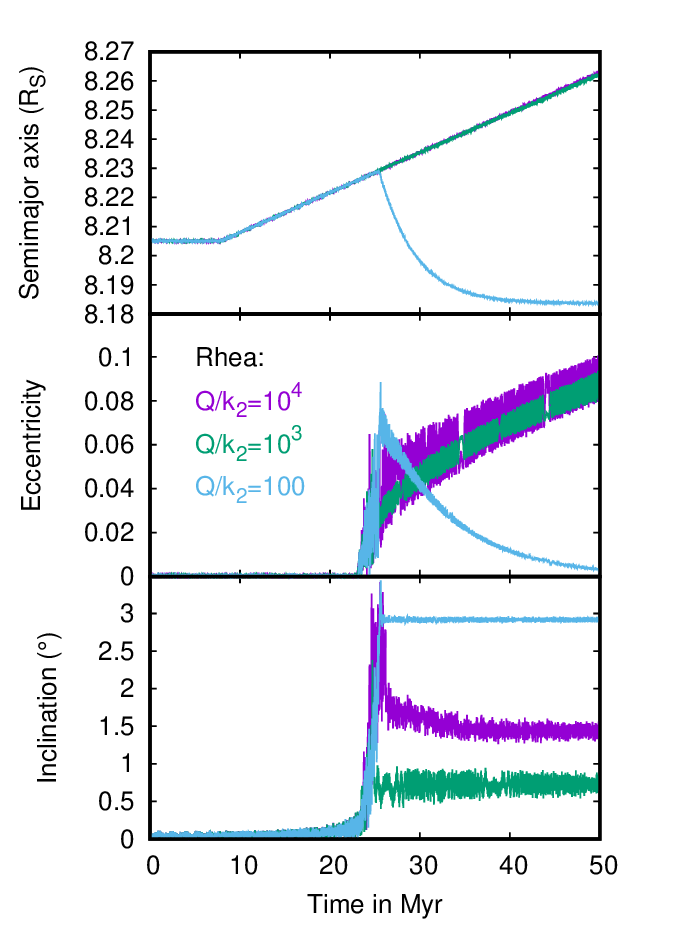}
\caption{Simulations of Rhea's encounter with the solar evection resonance using different parameters for the width of Saturn's resonant mode and Rhea's tidal dissipation. The simulations plotted on the left uses the current obliquity of Saturn, while those shown on the right use obliquity of $15^{\circ}$ for Saturn. From top to bottom, the panels plot Rhea's semimajor axis, eccentricity and inclination. The $Q/k_2$ values indicated on the plots (referring to lines with matching colors) are those used for Rhea, while the tides within Saturn were modeled using resonant lock tides. On the left, the green and the light blue lines had $\sigma=10^{-6}$ for the resonant mode, while the orange and the red lines had $\sigma=10^{-5}$. On the right, all simulations had $\sigma=10^{-6}$. The evolution timescale is $a/\dot{a}=6$~Gyr, as found for Rhea by \citet{lai20}. This simulation included the Sun, Jupiter, Saturn, Titan, and Rhea.\label{modes}}
\end{figure}

Fig. \ref{modes} shows a number of simulations of Rhea crossing the evection resonance using the integrator {\sc simpl} modified for resonance locking tides. Left-hand panels plot simulations that use an obliquity of 25$^{\circ}$ for Saturn, close to the current one; see the caption for tidal mode widths and Rhea's tidal properties. One of the simulations shows a long-term resonance capture (orange line), one ``drops out'' from the resonant mode (light blue line), while the other two simulations have Rhea staying with the mode but moving beyond the evection resonance (red and green lines). However, the bottom panel that plots inclination indicates that the evection and associated resonances also significantly excite the {\it inclination} of Rhea, well in excess of the observed value ($i=0.33^{\circ}$).

The excitation of inclination appears to be chaotic and occurs concurrently with the evection resonance. The main inclination resonance that is located very close to evection has an argument $\lambda_S-\varpi_S+\Omega-\Omega_S$ where $\Omega$ is Rhea's longitude of the ascending node and $\Omega_S$ that of Saturn's heliocentric orbit. As our reference plane is the Laplace plane of Rhea's orbit, which is close to Saturn's equatorial plane, $\Omega_S$ is equivalent to the longitude of Saturn's equinox. Both $\Omega_S$ and $\varpi_S$ are relatively slow-precessing (with periods of about 1.8~Myr and 50 kyr), so the resonance happens because $\dot{\lambda}_S = n_S \approx -\dot{\Omega}$. One can think of this resonant term as a combination of the Sun's ``annual equation'' \citep[a term in classical lunar theory;][]{bro61} and the Sun's main secular perturbation on the inclination of Rhea. In other words, the Sun induces perturbations in Rhea's inclination once per precession period, and solar perturbations vary in strength during one orbit of Saturn because of the planet's orbital eccentricity. When these two cycles are commensurable, we get this strong resonant perturbation. This perturbation is chaotic, as its strength depends on the eccentricity of Saturn's orbit, which varies in the $0.01 < e_S < 0.09$ range with a period of $5 \times 10^{4}$~yr due to Jupiter's perturbations. 

As the inclination-type resonance that is co-located with the evection includes the node between the equator and orbit of Saturn, we expect its strength to be proportional to Saturn's obliquity. In the right-hand panels of Fig. \ref{modes} we show simulations using an obliquity of $15^{\circ}$ for Saturn, and inclination changes that accompany evection resonance crossing do appear to scale with the obliquity. As we discussed in Section \ref{intro}, the obliquity of Saturn likely changed in the past couple of Gyr due to the spin orbit-resonance between Saturn's spin axis precession the Solar System secular mode $s_8$, and this resonance evolved due to Titan's orbital expansion \citep{sai21a}. In Figure \ref{aa_plot} we plot (using vertical blue lines) approximate semimajor axis of Titan for which the obliquity of Saturn was about $15^{\circ}$ and the semimajor axis of Titan at which the spin-orbit resonance was established (the line labeled $0^{\circ}$). Consistently fast past orbital evolution of Titan would therefore mean that the Saturnian system had only a small obliquity beyond about 1~Gyr ago\footnote{This is a rough minimum estimate that assumes the Saturn is still in the spin-orbit resonance. Estimates consistent with our scenario of instability of Proto-Hyperion would push the onset of spin orbit resonance closer to 2~Gyr ago}. Therefore there is a hypothetically possible case in which Rhea was not in resonance lock for much of its history, while Titan was, in which case Rhea crossed the evection resonance before Saturn had a significant obliquity (track "H" in Figure \ref{aa_plot}), and this possibility needs to be addressed.

\begin{figure}
\plottwo{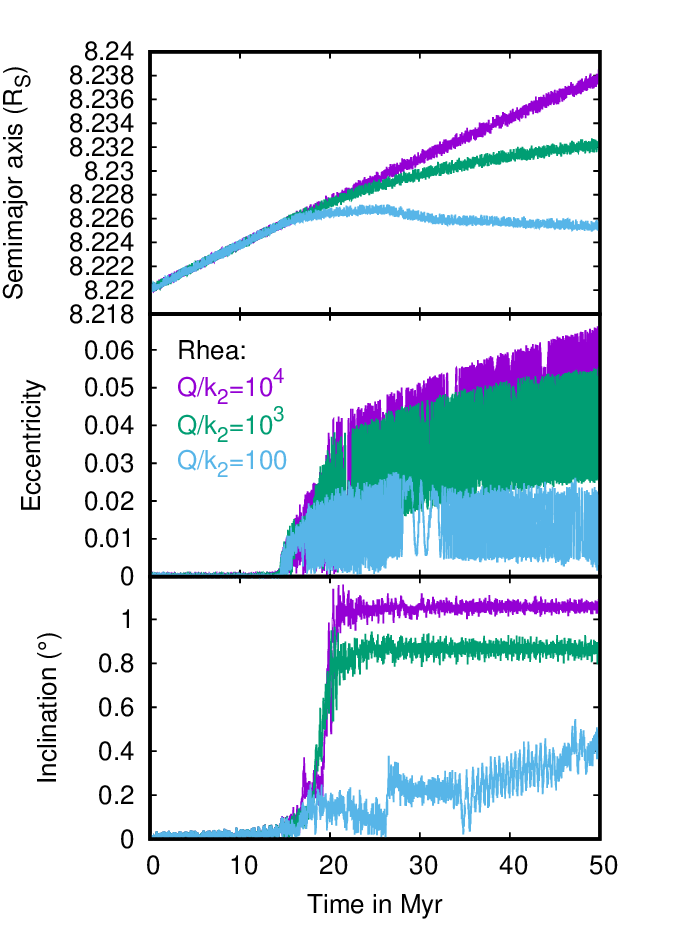}{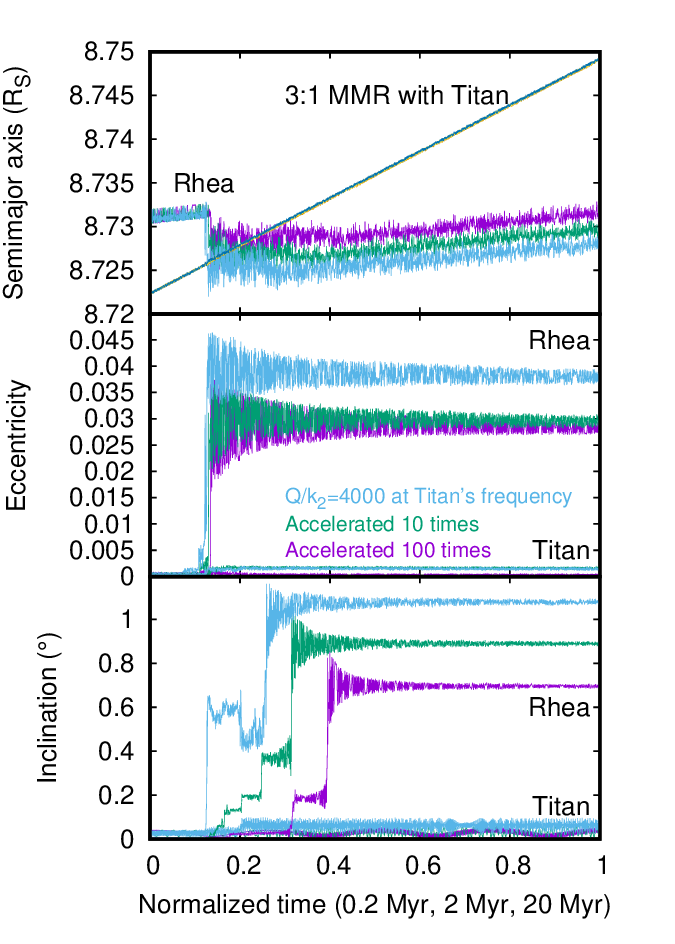}
\caption{Simulations exploring the hypothetical case in which Titan's migration was consistently fast but Rhea's was not. Left: Simulations of Rhea crossing the solid evection resonance assuming near-zero zero obliquity of Saturn. Right: simulations of Rhea and Titan divergently crossing their mutual 3:1 MMR.\label{31titan}}
\end{figure}

Figure \ref{31titan} explores this hypothetical scenario of ``fast Titan'' and ``slow Rhea'' being long-term conditions. The left hand side shows the simulations of Rhea crossing the evection resonance in a non-tilted Saturnian system. Note that even if Saturn's spin pole is perpendicular to the invariable plane of the Solar System, Saturn's orbital inclination induces effective obliquity of about a degree. We assumed $Q/k_2=4000$ for Saturn at Rhea's orbital frequency in order to be consistent with our hypothetical scenario of long-term slow migration of Rhea. While most simulations still produce the inclination in excess of the observed one (partly due to very slow resonance crossing), the simulation that assumes very dissipative Rhea produces an inclination consistent with observations. However, as shown in Fig. \ref{aa_plot}, our hypothetical scenario requires Rhea and Titan to cross their mutual 3:1 resonance. Right-hand panels in Fig. \ref{31titan} shows this crossing for a range of Titan's migration rates, and, even for an initially zero-inclination Rhea and Titan, this crossing gives Rhea an inclination well in excess of the observed one. Therefore we conclude that even this hypothetical scenario of slow Rhea and fast Titan, which is in conflict with the current migration rates \citep{lai20, jac22}, cannot avoid the conclusion that Rhea should have a significantly higher inclination due to its past orbital evolution.  

In some cases, the orbital inclination of a moon can be damped by obliquity tides \citep{chy89}, but we argue that this is not applicable to Rhea. Large-scale inclination damping requires large forced obliquities, either through Cassini State transition like the one experienced by the Moon \citep{che16}, or spin-orbit resonance, as proposed for exoplanets \citep{mil19}. The forced obliquity of Rhea is currently $\theta=0.03^{\circ}$ \citep{che14}, and the timescale for inclination damping is longer than that for eccentricity damping by a factor of $7 (\sin{\theta}/\sin{i})^2$ \citep{chy89}, which in the case of Rhea is on the order of $10^3$. If Rhea is as dissipative as ``Enceladus in equilibrium'' with $Q/k_2=100$, eccentricity damping scale is on the order of 10~Myr (cf. light blue lines Fig. \ref{modes}), which puts the timescale for inclination damping to many Gyr. Resonant tides excited in satellite oceans offer a different mechanism of inclination damping \citep{tyl08}, but \citet{che14} find that ocean obliquity tides would produce less dissipation in Rhea than classic obliquity tides that assume $Q/k_2 \simeq 100$. The above upper limits on dissipation may be overestimates, as Rhea currently does not appear to be in hydrostatic equilibrium \citep{tor16}, and a solid tidal Love number $k_2=0.01$ \citep{che14} may be more applicable. Therefore, we argue that the damping of multi-degree inclination of Rhea within the last 0.5 Gyr (or even 2~Gyr) is implausible. This conclusion is in not in conflict with the apparent fast damping of Titan's inclination \citep{dow25}, as Rhea not only has a much larger forced-obliquity-to-inclination ratio due to being closer to Saturn, but is also expected to have orders of magnitude higher rigidity (and lower tidal Love number $k_2$) than Titan \citep{che14}.

\section{Inner System Instability}\label{inner}

Due to multiple lines of evidence indicating that the inner moons of Saturn may be only up to 100~Myr old, \citet{cuk16} looked for a mechanism that could lead to a recent re-formation of the system. They noted that the re-accretion timescales for mid-sized moons are much shorter than their tidal evolution timescales, so it is reasonable to assume that the current inner moons were preceded by a previous generation of moons that had a similar total mass and angular momentum. The exception is the rings, which cannot accrete into the moons, and might have had little or no mass before the hypothesized disruption event. As the innermost moons (Enceladus, Mimas, Janus etc.) and especially the rings themselves are less massive than Tethys, Dione and Rhea, it is plausible that the former were a by-product of the re-accretion of the latter. In particular, a disruption collision between Dione- and Rhea-sized moons would involve a large majority of the mass of the inner system. Note that the density of Tethys is much lower than that of Dione and Rhea, indicating that Tethys does not have a sizable rocky core. Therefore there is a distinct possibility that Tethys was accreted mostly from mantle material, and that before the last re-accretion there were only two large inner moons, the cores of which were inherited by Dione and Rhea.

The dynamical mechanism for destabilizing precursors to Dione and Rhea and making them collide is not immediately clear. Mutual resonances are not strong enough to excite the eccentricities of Dione and Rhea beyond 0.1 required for orbit crossing and mutual collision \citep{cuk16}. However, as we saw in the previous Section, Rhea is relatively close to the evection resonance with the Sun, and \citet{cuk16} found that Rhea evolving though equilibrium tides with Saturn's $Q=1700$ \citep{lai12}  can reach high eccentricities. Additionally, this tidal $Q$ of Saturn roughly matches what is necessary for Rhea to evolve from orbits much closer to Saturn to the evection resonance. When a second, Dione-like, moon is introduced into the simulation and allowed to enter a mean-motion resonance with the Rhea-like moon that is participating in the evection resonance, chaotic interaction leads to orbit crossing and collisions. While in Section \ref{rhea} we did not include any additional inner moons, it is clear from Fig.\ref{modes} that Rhea entering the evection resonance due to resonance lock tides can also reach high eccentricities, as long as its satellite tides are not too strong.

\citet{hyo17} were the first to simulate numerically the collision between moons similar to Dione and Rhea, as proposed by \citet{cuk16}. They found that the initial collision significantly disrupts the two moons and produces large-scale debris. However, \citet{hyo17} concluded that such a collision cannot produce Saturn's rings as almost all debris is re-accreted on the two large moons. This finding was surprising, as the initial moons had the largest escape velocities, and their remnants would be even less able to grow rather than being eroded by the collisions at similar velocities. While \citet{hyo17} used SPH modeling to simulate the initial collision, they used an N-body code to follow the fragments, and in that N-body code the coefficient of restitution for the normal component of collisional velocities was only 0.1 \citep{hyo17}. Therefore, in a purely head-on collision, only $1\%$ of the initial kinetic velocity would be taken into account when deciding whether there is accretion or not \citep[][ Eqs. 3-4]{hyo17}. This approach all but guarantees re-accretion and cannot be considered valid.  

\citet{teo23} re-visited the collision between Dione- and Rhea-like moons with a similar setup and reached a different conclusion, that such a collision could deliver a significant amount of material to within the Roche limit. The initial collision in the work of \citet{hyo17} produced debris with a smaller range of semimajor axes, and without having the pericenter within the Roche limit of Saturn. While \citet{teo23} had a much larger resolution in their SPH simulations, the main reason for this difference may be orientation of the collision in the Saturn-centric frame, with \citet{teo23} finding that largely tangential collisions (which are to be expected for barely-crossing orbits) produce the largest spread in semimajor axis \citep[the orientation of the collision in ][ is not specified]{hyo17}. In any case, we take the results of \citet{teo23} as  a proof that a collision between two moons roughly the size of Dione and Rhea can produce widespread debris in the inner system, which can disrupt further smaller moons (if there were any) and deliver material to the rings.
 
While an inner system instability due to the evection resonance is still possible in the context of resonance lock tides and rapid migration of Rhea and Titan\citep{ful16, lai20}, this outcome is not assured as evection resonance can be crossed leaving Rhea (or its progenitor) with sizable but not destabilizing eccentricities (Fig. \ref{modes}). The idea of a recent cataclysm in the outer system, originally proposed by \citet{ham13} and \citet{asp13} and now developed in more detail by \citet{wis22} and the present work offers new dynamics pathways to the re-accretion of the inner system. If the outer system instability happened about 500~Myr ago due to destabilization of proto-Hyperion due to rapid migration of Titan, it appears unlikely that a subsequent inner system instability had a completely independent origin.     

\citet{cuk16} did not propose resonances with Titan as a possible source of instability because they assumed that the instability originated in the inner system. It has long been suspected that the substantial eccentricity of Titan must have been generated recently, as tidal dissipation would have otherwise damped it on the age of the Solar System \citep{sag82}. Therefore, \citet{cuk16} assumed that before the inner mid-sized moons acquired excited orbits, Titan had a nearly-circular orbit that was excited due to the inner system instability. They proposed that Titan's perturbations may prevent a ring of debris close to Titan's 3:1 MMR from accreting, and that the interaction between the two (with short-lived smaller moons as intermediaries) may have excited Titan's eccentricity. However, if Titan's orbital excitation predates the inner system cataclysm, not only does Titan's eccentricity not need to be excited by the inner moons, but it can be the source of their instability. Currently Rhea is between Titan's 4:1 and 3:1 mean-motion resonances, but the latter is not a viable candidate as it overlaps with the 4:1 MMR with Hyperion and \citet{cuk16} have found that a mid-sized moon in Titan's 3:1 resonance invariably destabilized Hyperion. Therefore we concentrate on the inner 4:1 MMR with Titan.

\begin{figure}
\plottwo{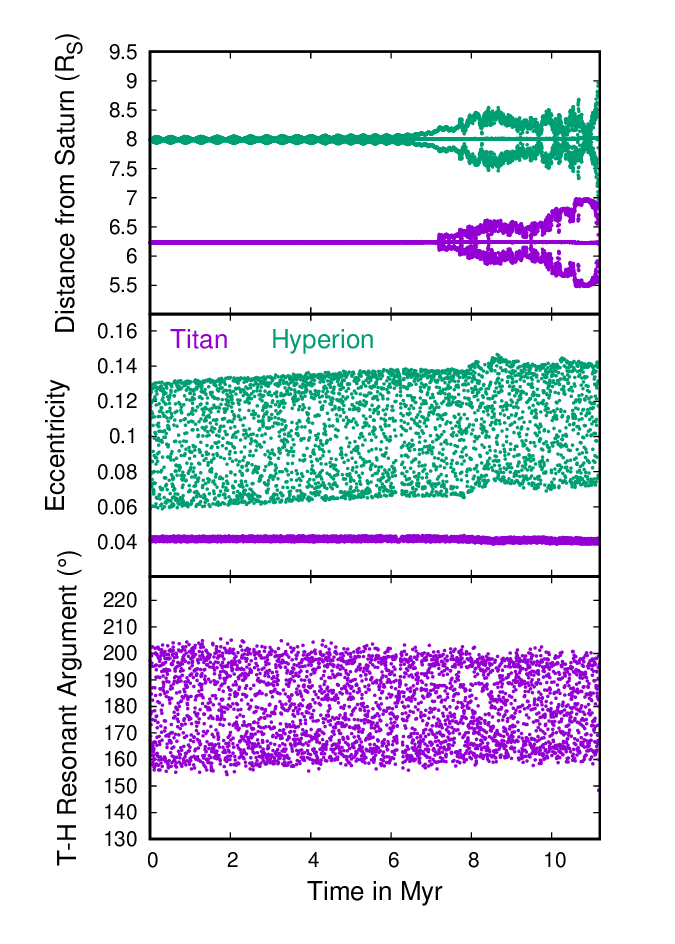}{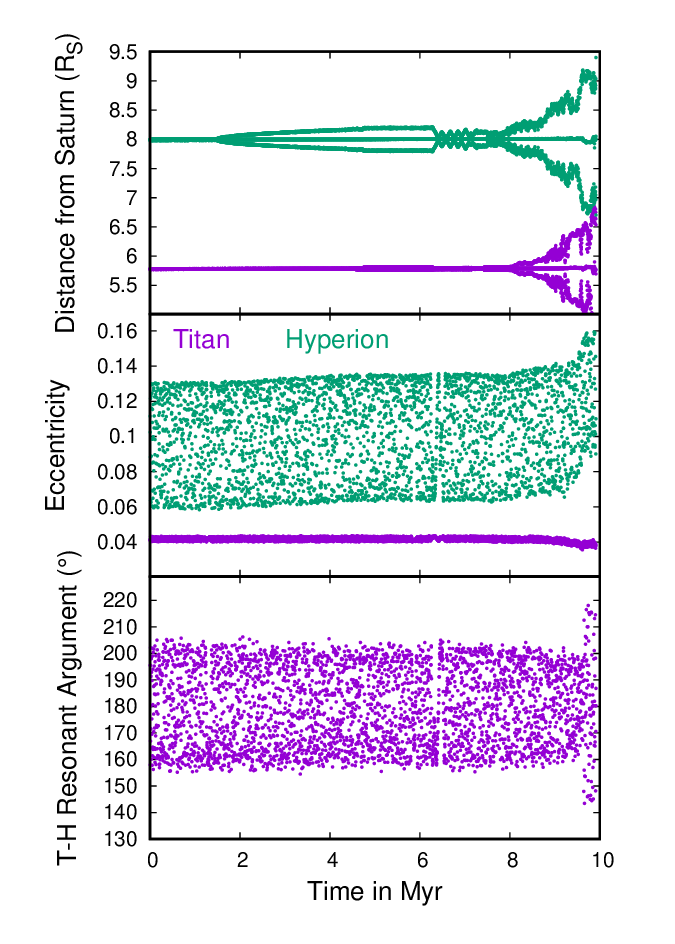}
\caption{Simulations of the destabilization of the inner system through 4:1 resonance with eccentric Titan. In these simulations the outer moon evolves at the similar rate to present-day Rhea and catches up with Titan's resonance. The top panel shows the pericenter, semimajor axis and apocenter of two hypothetical inner moons, the outer of which enters the 4:1 MMR with Titan. The middle panel shows the eccentricities of Titan and Hyperion, and the bottom panel shows the Titan-Hyperion 4:3 MMR resonant argument.\label{fast_rhea}}
\end{figure}

In order to test the 4:1 MMR with Titan as the source of the inner system instability, we set up simulations with two hypothetical past moons in the inner system. The total mass of the two moons was $2.1 \times 10^{-9}$~$M_{Sun}$, very close to the total mass of all the moons out to Rhea (plus the rings). The outer of the two moons was placed just interior of the 4:1 MMR with Titan, around 8~$R_S$. We put the two satellites on mutually non-resonant orbits, with one configuration having them closer together than their 3:2 MMR (left hand side of Fig. \ref{fast_rhea}) and the other having the two moons wide of their 3:2 MMR (right-hand side in Fig. \ref{fast_rhea}). To approximately match the inner system angular momentum, the two moons had masses $m_1=0.8 \times 10^{-9} M_{Sun}$ and $m_2=1.3 \times 10^{-9} M_{Sun}$ (in the more compact configuration) and $m_1=0.7 \times 10^{-9} M_{Sun}$ and $m_2=1.4 \times 10^{-9} M_{Sun}$ (for the more separated pair), where indices 1 and 2 refer to the inner and outer moon, respectively. The two inner moons were given orbits that were close to circular and planar, while Titan was assumed to have $e=0.04$ and $i=1^{\circ}$ at this time (we found Titan's inclination to have little effect on the outcome). The outer moon experienced $Q=330$ for Saturn, similar to the present-day Rhea, while the inner moon and Titan experienced $Q=2000$ and $Q=100$ for Saturn, respectively (i.e. the innermost moon was not in resonance lock unlike the other two). Just like Rhea is currently catching up with Titan now, the outer hypothetical moon was catching up with Titan's resonances, with the possibility of capture.

Figure \ref{fast_rhea} shows that in both of the simulations of ``proto-Rhea'' catching up with Titan's 4:1 resonance the inner moons experience instability very soon after the resonant encounter. While only Proto-Rhea is affected by the resonance with Titan, both moons become eccentric as mutual perturbations spread excitation to the inner moon (``Proto-Dione''). Eventually the two inner moons acquire crossing orbits, at which point we stop the simulation using {\sc simpl}. Such moons will typically collide on $10^3$~yr timescales \citep{cuk16}. In one simulation there was barely any effect on Hyperion and the Titan-Hyperion resonance, while in the other Hyperion does become somewhat more eccentric and the amplitude of the Titan-Hyperion 4:3 MMR does become wider, potentially affecting somewhat our estimate of Hyperion's age but not disrupting the resonance in a manner inconsistent with the present system. 

\begin{figure}
\plotone{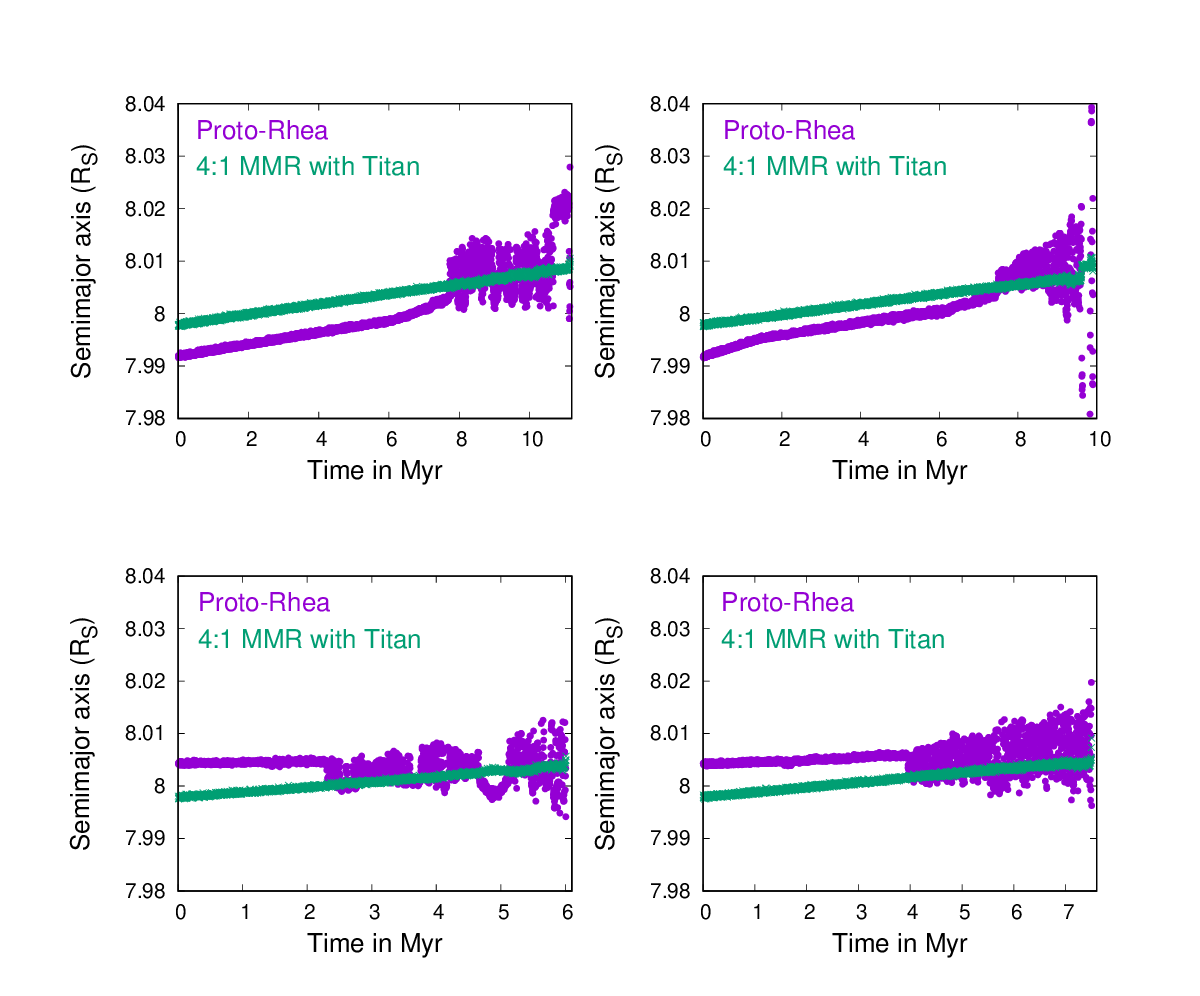}
\caption{Semimajor axes of Proto-Rhea and the 4:1 inner resonance with Titan in the four simulations shown in Fig. \ref{fast_rhea} (corresponding to the top two panels in this figure), and \ref{slow_rhea} (bottom panels). \label{sma}}
\end{figure}

\begin{figure}
\plottwo{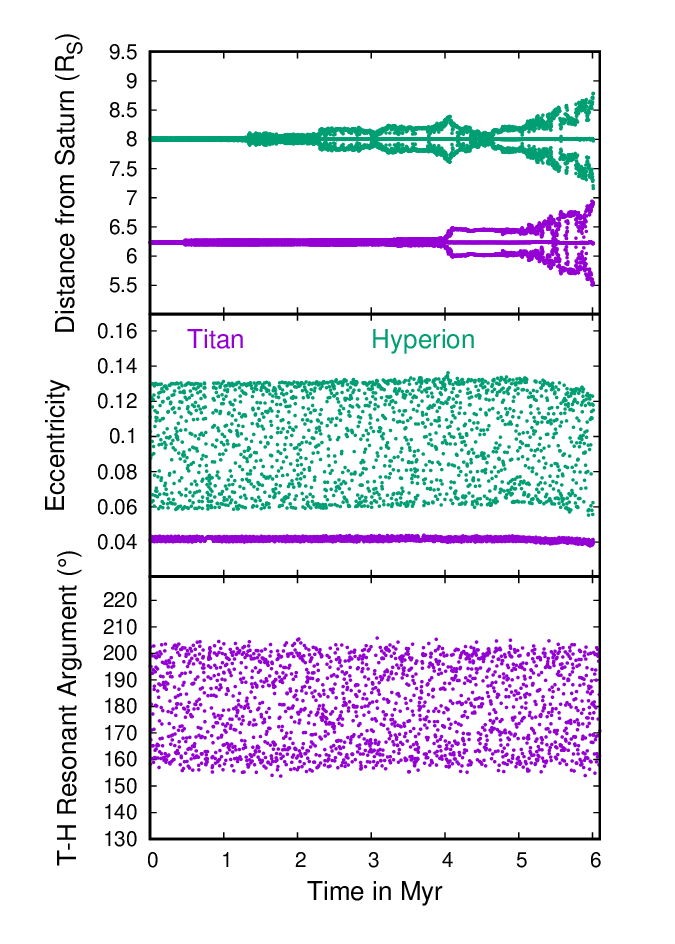}{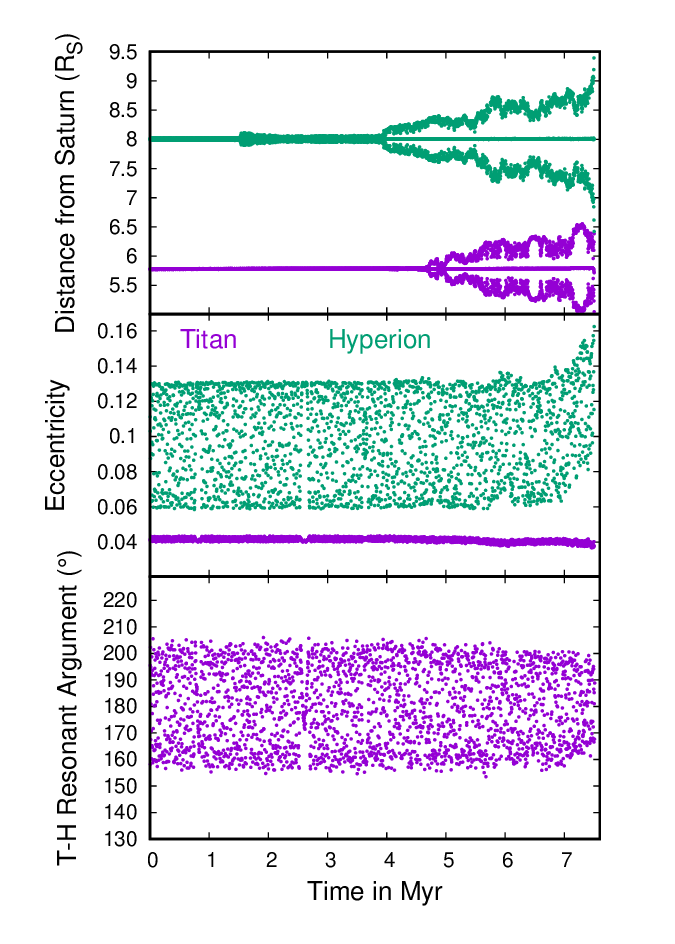}
\caption{Simulations of the destabilization of the inner system through 4:1 resonance with eccentric Titan. In these simulations the outer moon is not locked to a resonant mode and Titan's 4:1 MMR catches up with this moon. The top panel shows the pericenter, semimajor axis and apocenter of two hypothetical inner moons, the outer of which enters the 4:1 MMR with Titan. The middle panel shows the eccentricities of Titan and Hyperion, and the bottom panel shows the Titan-Hyperion 4:3 MMR resonant argument. \label{slow_rhea}}
\end{figure}

The top two panels in Fig. \ref{sma} show the semimajor axis evolution of Proto-Rhea in the simulation from Fig. \ref{fast_rhea}. Variations in the tidal migration rate of Proto-Rhea are due to temporary captures into the sub-resonances of the 4:1 MMR with Titan that involve only the eccentricities and/or inclinations of the inner moons. We find that these captures are always short-lived, and eventually Proto-Rhea reaches the core of the resonance that includes sub-resonances involving Titan's eccentricity, at which point the resonance becomes chaotic. In Fig. \ref{fast_rhea} we assumed that Proto-Rhea was affected by a resonant mode within Saturn, in analogy to present day Rhea. However, the orbits of Rhea and Proto-Rhea are not identical, and resonant modes also shift over time, so we cannot be sure that Proto-Rhea was also resonance locked and on an orbit converging with Titan's. Resonant encounters depend profoundly on whether the orbits of the two bodies are converging or diverging, so we need to explore the possibility that Proto-Rhea and Titan encountered their 4:1 MMR on diverging orbits. The bottom two panels in Fig. \ref{sma} plot simulations identical to the top panels, except that Proto-Rhea was placed outside the 4:1 MMR with Titan and was experiencing Saturn's tides with $Q=2000$, which made it migrate slower than Titan. Once Titan's resonances catch up with Proto-Rhea chaotic interaction ensues, resulting in orbit-crossing between the two inner moons. Fig. \ref{slow_rhea} plots the eccentricities of the moons involved and the Titan-Hyperion resonant argument in these two simulations.  

Figure \ref{slow_rhea} shows, similar to Fig. \ref{fast_rhea}, that simulations with initially less compact inner satellite pairs (right hand panels) tend to produce more back-reaction on Titan and Hyperion than more compact inner systems (left-hand side). This is understandable as the inner moons must reach higher eccentricities for their orbits to cross in the less compact configurations. In the less compact configuration explored in Fig. \ref{slow_rhea} the eccentricity of Hyperion increases significantly while the Titan-Hyperion resonance amplitude tightens, potentially interfering with the estimates of the resonance age, but not disrupting the resonance In any case, the preliminary simulations of two non-resonant inner moons seem to suggest that they always become unstable and collide if the outer satellite enters the 4:1 MMR with Titan (on an orbit with $e_T=0.04$), regardless of the outer moon's migration rate. In the handful of preliminary simulations we have completed, the Titan-Hyperion resonance is not disrupted, and Titan's eccentricity is not dramatically affected by the resonance with the inner moon.  

\begin{figure}
\plotone{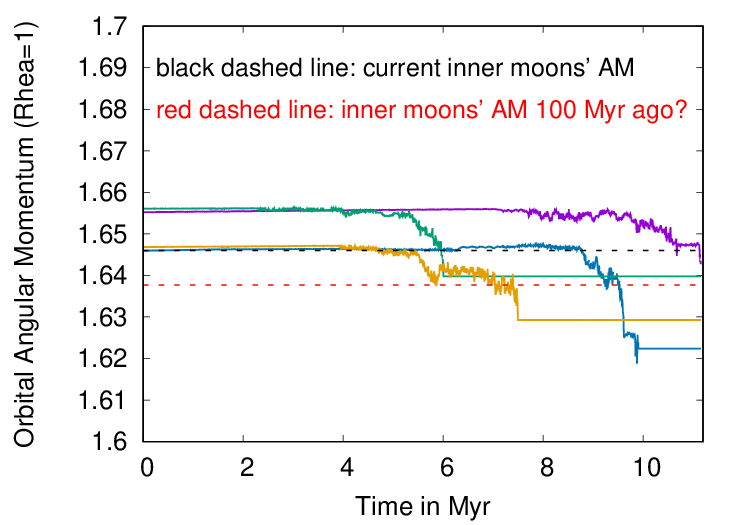}
\caption{Angular momentum of the two inner moons in the simulations plotted in Figs. \ref{fast_rhea} (purple and blue lines) and \ref{slow_rhea} (green and orange lines) in units of the present orbital angular momentum of Rhea. Total angular momentum of inner moons out to Rhea is plotted as a black dashed line, and the red dashed line shows angular momentum 100 Myr ago, assuming present orbital migration rate of Rhea in the past. \label{angmom}}
\end{figure}

It is expected that, once the inner moons collide, some orbital energy would be lost to heat during impacts, but total mass and orbital momentum would be conserved \citep[apart from a small fraction of debris that crosses Titan's orbit and will likely be removed][]{teo23}. While the mass of our two moons is chosen to match the present mass of the inner system, we can ask how well the angular momentum fits the current system. Figure \ref{angmom} plots the angular momentum of the inner two moons, calculated as:
\begin{equation}
L= \frac{m_1}{m_{Rhea}} \sqrt{\frac{a_1}{a_{Rhea}} (1-e_1^2)} \cos{i_1} + \frac{m_2}{m_{Rhea}} \sqrt{\frac{a_2}{a_{Rhea}} (1-e_2^2)} \cos{i_2} \label{am_eq}   
\end{equation}
It is clear that in these units present-day Rhea, with its low eccentricity and relatively low inclination, has $L \approx 1$. The solid lines plot thus calculated angular momentum of the two moons in simulations from Fig. \ref{fast_rhea} (purple and blue lines) and Fig. \ref{slow_rhea} (green and orange lines). The black dashed line plots the current angular momentum of the inner system (Rhea and the moons interior to it). Initially, our more compact configurations had angular momentum slightly above the current, while the less compact configurations had almost exactly the current angular momentum. When the two inner moons' orbits cross at the end of the simulations, all four simulations have angular momenta somewhat below the current one, mostly due to their significant eccentricities ultimately induced by Titan. However, as the angular momentum decreased only by 1-2\%., it is clear that small tweaks to our initial conditions could produce simulations that give us (on average, given that they are stochastic) the current angular momentum of the inner system. Additionally, angular momentum at the time of the inner system's re-accretion is not known. If we assume Rhea migrated at the present rate (i.e. it was resonance locked) over the past 100~Myr, angular momentum would be the one plotted by the red dashed line, which is now lower than two of our end-states. Therefore, our preliminary results indicate that the two colliding moons discussed in this Section can match the angular momentum of the present inner system. 

Our scenario presented here is somewhat similar to that proposed y \citet{cuk16} and subsequently studied by \citet{hyo17} and \citet{teo23}, with the difference that the instability is now caused by the 4:1 MMR with Titan rather than the evection resonance with the Sun. Therefore we must compare the collisional velocities we get in these simulations to those used by collision studies using SPH simulations. To get the collision velocities, we took the last non-crossing orbit in the output (output interval of 10$^3$~yr is comparable to the collision timescale) and, if necessary, increased the eccentricities slightly for the inner moon's apocenter to match the outer moon's pericenter. 

\begin{deluxetable}{cccc}
\tablenum{1}
\tablecaption{Collision distances and velocity at infinity for the four simulations plotted in this Section. We assumed that collision happens when the ascending node of one moon matches the descending node of the other. \label{impact}}
\tablehead{\colhead{Configuration} & \colhead{Proto-Rhea and Titan} & \colhead{Collision distance [$R_S$]} & \colhead{Collision $v_{\infty}$[km s$^{-1}$]}}
\startdata 
compact & converging & 6.93  &  1.69 \\
wide  & converging & 6.62 & 2.36\\
compact & diverging & 7.04  &  2.05 \\
wide  & diverging & 6.40 & 2.22\\
\enddata
\end{deluxetable}

\citet{teo23} used $v_{\infty}=2$~km~s$^{-1}$ and $v_{\infty}=3$~km~s$^{-1}$ in their successful simulations of inner system disruption that produced mass within the Roche limit. However, the faster impacts delivered more material to the potential rings, and also produced large-scale disruption of the colliding bodies for a wider range of impact angles. The impacts listed in Table \ref{impact} have velocities closer to 2~km~s$^{-1}$, with three of the collisions being faster and one slower. In general, wider configurations lead to larger collision velocities, as larger eccentricities are necessary for orbit-crossing and are usually accompanied by higher inclinations. Like \citet{teo23}, we assumed that the moons are colliding at the apocenter of one and the pericenter of the other, and also at the ascending node of one and the descending node of the other. The assumption about opposite nodes is essential for reaching large impact $v_{\infty}$, but is not very restrictive, as two barely-touching orbits must collide close to the equatorial plane of Saturn, with either aligned or anti-aligned nodes. We also note that, among the simulations done by \citet{teo23}, the most destructive ones are not the best match for the present system, as the cores of Proto-Rhea and Proto-Dione need to stay largely intact given that Tethys appears to be made solely from mantle material. 

There is clearly a need for future work on the detailed mechanism 4:1 resonance with Titan can lead to inner system re-accretion and ring formation. While the need to have the outer moon at 4:1 MMR with Titan makes our system more compact than those modeled by \citet{teo23}, our impact distance is also closer to the planet and the Roche limit. Our moons are a bit more massive, but also our impactor is a larger fraction of the target than in \citet{teo23}, who used the current masses of Dione and Rhea. As we conducted only a few simulations, a study of destabilization with a wider range of parameters (notably different positions and masses of the inner moon) is necessary to fully explore the possibilities. Simulations of bodies on crossing orbits like the ones in Section \ref{outer} can obtain more realistic impact conditions, to be followed by more extensive SPH studies of the collisions themselves along the lines of \citet{teo23}. 

\section{Titan-Iapetus 5:1 Resonance}\label{iapetus}

In Section \ref{outer} we have shown that the outer system instability involving an additional moon, Proto-Hyperion, in the outer 2:1 MMR with Titan produces dynamical excitation of Iapetus's orbit that can in principle explain Iapetus's current orbital inclination. However, we cannot directly compare the outcomes in Fig. \ref{all_p-h4.4} with the orbit of Iapetus and Titan and Iapetus should since have crossed their mutual 5:1 resonance. Assuming Titan's rapid migration found by \citet{lai20}, this resonance was crossed less than 50 Myr ago \citep{pol18}.

In order to study the effects of the Titan-Iapetus 5:1 MMR we conducted follow-up simulations using {\sc ssimpl}, starting with the outcomes plotted in Fig. \ref{all_p-h4.4}. Each of the 42 ``successful'' outcomes, in which Titan and Proto-Hyperion collided and Iapetus was retained, was used to set up five simulations of Titan-Hyperion 5:1 MMR crossing. The only elements carried over from the original simulations were the free inclinations and eccentricities of Iapetus. As we aim to study whether these evolutionary paths could result in the present orbit of Iapetus, we use the current semimajor axis for Iapetus. Our thinking is that, while there are physical reasons for starting with certain (almost zero) eccentricities and inclinations for Iapetus in Section \ref{outer}, the initial semimajor axis was unknown and therefore we can only say that it has the present value post-instability. Similarly, for Titan we use a semimajor axis just short of 5:1 resonance with Iapetus, as well as almost-current $e=0.03$ and the current inclination $i=0.3^{\circ}$. As this resonance happened a few tens of Myr ago, the orbit of Titan was very unlikely to have been very different. The five simulations for every initial orbit of Iapetus differed by slightly offset initial semimajor axes of Titan. We used $Q=100$ for Saturn as experienced by Titan and ignored tidal migration of Iapetus.

\begin{figure}
\plottwo{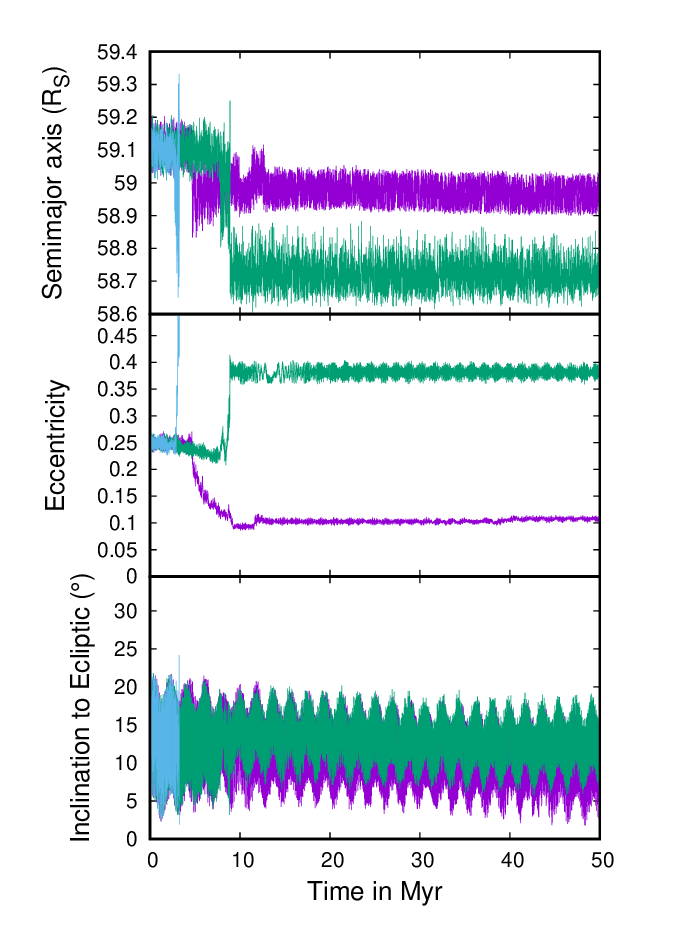}{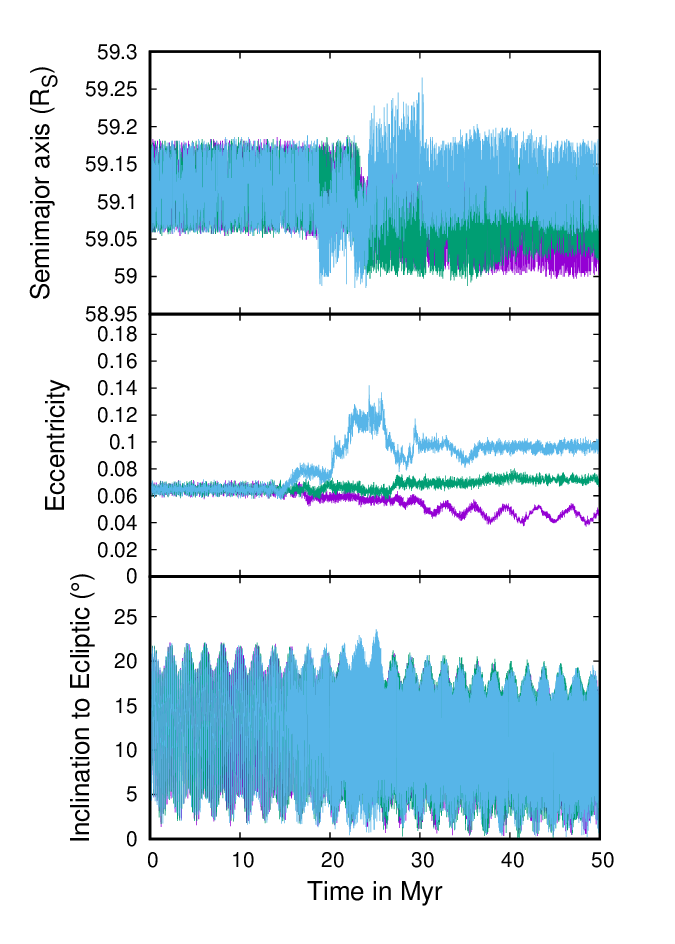}
\caption{Simulations of the Titan-Iapetus 5:1 mean-motion resonance crossing for three clones each of two initial conditions based on the end-states of simulations of outer system instability shown in Fig. \ref{all_p-h4.4}. The end-states of the five stable clones plotted here are indicated in Fig. \ref{iap_all} by red circles. Inclination is plotted to ecliptic, so free inclination is roughly equal to the semi-amplitude of the inclination variations in the bottom panels.\label{iap_plots}}
\end{figure}

Figure \ref{iap_plots} plots three simulations each for two very different initial conditions for Iapetus. On the left-hand side we plot the outcomes for a high-eccentricity ($e_I=0.25$) initial orbit of Iapetus. Three of the clones of Iapetus ended up being ejected, as shown by the orbit plotted in light blue. The orbit plotted in green stayed bound, but with a very high eccentricity, while the purple orbit ended up with a much smaller eccentricity. On the right-hand side we see orbits with relatively small initial eccentricity ($e_I=0.06$) which then diverge when the resonance with Titan is encountered. Notably, the divergence in eccentricity due to the 5:1 MMR with Titan is much larger than that in inclination, indicating that the large inclination of Iapetus must have been mostly generated during the outer system instability discussed in Section \ref{outer}. On the right-hand side the simulation plotted in purple shows eccentricity oscillations indicative of the secular resonance with the planets described by \citet{cuk18}. Notably, this is also the orbit most similar to the actual orbit of Iapetus.  

\begin{figure}
\plotone{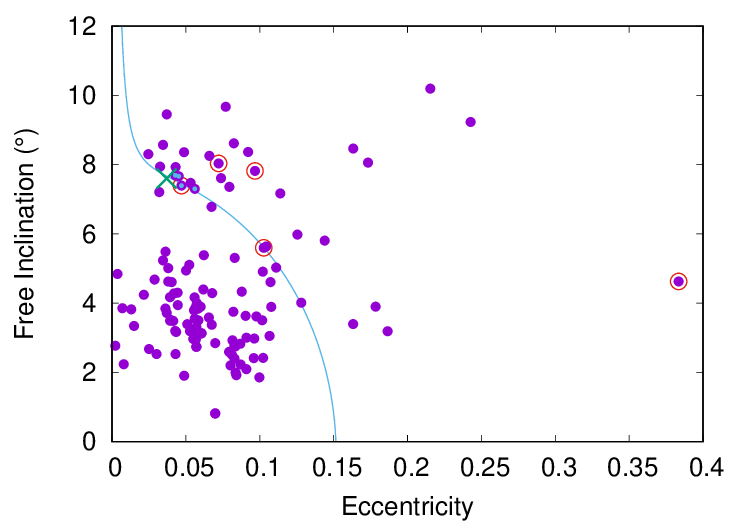}
\caption{Final eccentricities and free inclinations of Iapetus in 121 simulations (out of initial 210) in which Iapetus survived the crossing of its 5:1 MMR with Titan (purple dots). Current orbit of Iapetus is indicated by the green cross. The location of Iapetus's secular resonance with the giant planets \citep[from ][]{cuk18} is plotted in light blue, and light blue dots indicate the four simulations that are strongly affected by this resonance. Red circles indicate simulations plotted in Fig. \ref{iap_plots}. \label{iap_all}}
\end{figure}

Figure \ref{iap_all} plots the final eccentricities and free inclinations for 121 surviving clones (out of 210) of Iapetus in our simulations of the Titan-Iapetus 5:1 resonance. While the distribution of inclinations has not changed substantially, many fewer objects are at high eccentricities (21/121 or 17\% with $e_I>0.1$) compared to the top-left panel of Fig. \ref{all_p-h4.4} (16/42 or 38\% with $e_I>0.1$). So the net effect of the 5:1 MMR crossing with Titan is to depopulate high-e orbits and skew the remaining clones toward lower eccentricities. While the current orbit of Iapetus is on the lower-$e$ and higher-$i$ side of the distribution, the whole distribution of simulated outcomes is now more concentrated at low eccentricities than it was immediately after the instability. Therefore Iapetus's low eccentricity and high inclination are partly stochastic and partly due to the surviving clones being selected for low eccentricity. 

It is interesting that the overall survival rate of Iapetus's clones in our scenario (outer system instability followed by 5:1 MMR) is 121/300 or 40\%, but most of the losses happen because of the 5:1 resonance crossing with an eccentric Titan, which is practically unavoidable if Titan's migration is fast \citep{pol18}. It is fair to say that Iapetus had a close brush with ejection, in which case we would have little or no evidence it ever existed.   This should be taken as another clue that recent evolution of the Saturnian system was anything but staid and uneventful.

Blue dots in Fig. \ref{iap_all} indicate the clones that have large oscillations in eccentricity caused by the secular resonance with the planets \citep{cuk18}, and this includes one of the clones plotted in purple in the right-hand side of Fig. \ref{iap_plots} (clone with both the red circle and the blue dot). Since all of these clones have orbits very similar to that of real Iapetus (green ``X''), it appears that Iapetus just happens to be in orbital element space where the resonance has significant effects. In Fig \ref{iap_all} we also plot the location of the secular resonance as calculated by \citet{cuk18} with a blue line. There is no obvious correlation between the location of the secular in $e-i$ space and the orbital distribution of surviving clones. The fact that only low-$e$ orbits appear significantly affected by the secular resonance is likely due to the resonance being of first order in $e_I$ \citep[the resonant argument is $\varpi_I + \Omega_I - \varpi_{g5} - \Omega_{Eq}$, where $g_5$ refers to the solar system mode and ``Eq'' to Saturn's equinox;][]{cuk18}. Therefore variations in eccentricity caused by the secular resonance are much more noticeable if the eccentricity is low. At higher eccentricities the secular resonance only seems to cause a minor ``kick'' (as seen at 39~Myr in the purple curve in the left-hand side of Fig. \ref{iap_plots}). We must conclude that the current proximity between this secular resonance and Iapetus is simply due to chance and the resonance has had little or no effect on Iapetus's dynamical evolution.

The overall distribution of outcomes in Fig. \ref{iap_all} shows that the current orbit of Iapetus is well explained by an outer system instability followed by the 5:1 MMR with Titan. In particular, the probability of Iapetus obtaining a high inclination through this scenario is substantial (Fig. \ref{iap_all}), while it is a very unlikely ($<$1\%) outcome of Titan-Iapetus 5:1 MMR crossing, starting with Iapetus on an orbit with zero eccentricity and zero free inclination \citep{pol18}. We conclude that our proposed mechanism can fully explain the current orbit of Iapetus, and that it implies that Iapetus's high inclination developed relatively recently, about 400 Myr ago.

\section{Conclusions}\label{conclusions}

In this paper we present a new comprehensive model for the recent dynamical evolution of the Saturnian satellite system. It is motivated by the apparent youth of Saturn's rings \citep{ies19, est23}, apparent dynamical youth of the inner mid-sized moons \citep{cuk23, lai24}, the observed fast tidal migration of Titan \citep{lai20}, fast damping of Titan's inclination and eccentricity \citep{dow25}, and the fact that Saturn seems to have exited recently its assumed past spin-orbit resonance with the planets \citep{wis22}. We also note that the orbits of Hyperion (Section \ref{hyp}) and Rhea (Section \ref{rhea}) are not consistent with them being primordial. In particular, Hyperion appears to be only 400-500 Myr old, requiring a recent formation mechanism.

\begin{figure}
\plotone{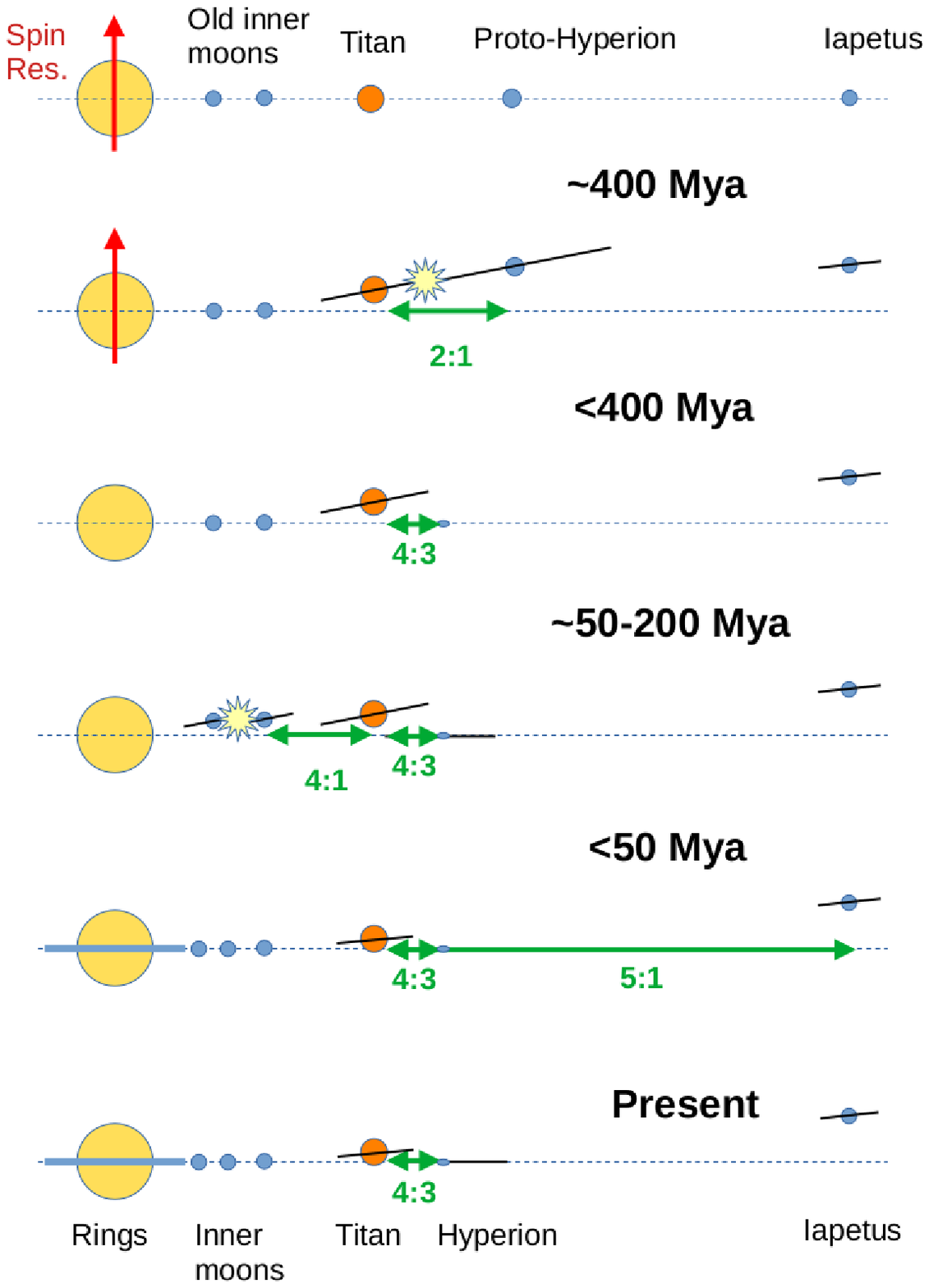}
\caption{A cartoon explaining the sequence of events proposed in this paper. The distances of the satellites from the planet represent their semimajor axes (not to scale). Inclinations are represented by the height above the dashed line (representing the Laplace Plane) and eccentricities are indicated by a radial solid line centered on the moon (connecting the pericenter and the apocenter). Green arrows are sued to indicate mean-motion resonances. The red arrow indicates that Saturn's spin is in resonance with the planetary secular mode, and many-pointed stars indicate collisions.}
\label{cartoon}
\end{figure}

We propose the following sequence of events, also illustrated in Fig. \ref{cartoon}:

1. An outer system instability occurred about 400~Myr ago, when Titan captured an exterior moon (Proto-Hyperion) into 2:1 mean-motion resonance, which eventually resulted in the collision between Proto-Hyperion and Titan. Hyperion was accreted from a small fraction of debris from this collision, to be later captured into its present 4:3 resonance with Titan. Pre-collision perturbations by Proto-Hyperion excited the eccentricities and inclinations of Titan and Iapetus (Section \ref{outer}).

2. Much later (50-200 Myr ago), the outer member of the pair of inner moons (``Proto-Dione'' and ``Proto-Rhea'') was caught into the 4:1 inner mean-motion resonance with eccentric and inclined Titan. The resulting orbital excitation of the inner moon's orbits led to their mutual collision (Section \ref{inner}). This collision generated the current inner moons and the rings \citep[cf.][]{cuk16, teo23}.

3. Less than 50~Myr ago Titan and Iapetus crossed their mutual 5:1 mean-motion resonance, which further scrambled the orbit of Iapetus and presented a significant risk of ejecting Iapetus (Section \ref{iapetus}).  


While the events described here took place hundreds of millions of years ago and are difficult to confirm directly, recent observations have consistently challenged previous models and revealed new dynamical pathways. Our hypothesis predicts a dynamically active and relatively young Saturnian system whose present configuration is the product of recent, dramatic events.

Future orbital, geophysical, and geological data, particularly from missions targeting Saturn’s moons, will provide essential tests of this scenario. These data would include independent determinations of Saturn’s axial precession rate and Titan’s orbital evolution, tidal and rotational parameters of Titan, and surface age of Titan and other moons.
 Whether or not our specific sequence of events is confirmed, we think this work helps frame new hypotheses about the evolution of Saturn’s satellite system. 

\begin{acknowledgments}
We wish to thank Jack Wisdom and Francis Nimmo for providing us additional information on their past work. M\'C also thanks Max Goldberg and Kevin Zahnle for useful discussions. Comments from two anonymous reviewers greatly improved the manuscript. M\'C was supported by NASA Solar System Workings program award 80NSSC22K0979.
\end{acknowledgments}

\begin{contribution}

M\'C set up, ran and analyzed numerical simulations. All authors helped write the paper.

\end{contribution}

%



\appendix

\section{Alternate Hypotheses on Hyperion's Orbital Evolution}\label{app}

Recently, \citet{gol24a} put forward a different hypothesis on the orbital evolution of Hyperion, and proposed that Hyperion is in (or close to) long-term equilibrium in which its eccentricity stays constant and resonant excitation by Titan is balanced out by tidal damping within Hyperion. However, this conclusion was reached by incorrectly combining the equations describing tidal dissipation and energy damping through non-principal axis rotation (equivalent to our Eq. \ref{edot} and Eq. \ref{sharma}) in Eq. 18 of \citet{gol24a}. The boost in dissipation is achieved by assuming that tidal dissipation rate can depend on spin frequency to the fourth power, as NPA rotation damping does. This way, constant 4$\times$ synchronous rotation would enhance {\it tidal} damping by $4^4$ times, which is combined with another erroneous factor of 100 from a spurious $e^2$ term to gain four orders of magnitude. If we assume constant 4$\times$ synchronous rotation, we could indeed increase the NPA rotation energy damping rate by an order of magnitude in Eq. \ref{sharma} (we originally used twice the synchronous rotation) but that rate is still more than two orders of magnitude too small to put Hyperion into equilibrium eccentricity state. To derive the correct NPA rotation damping rate one must use a consistent approach such as the one that produced Eq. \ref{sharma}, and not {\it ad hoc} mixing of tidal and rotational dissipation. Saturn's tidal forces are actually less relevant (compared to NPA rotation dissipation) if Hyperion is indeed in quasi-regular state in which the fastest rotation is along the long axis as proposed by \citet{gol24b}. As \citet{gol24b} argues, dissipation within this state is lower than that happening during more chaotic rotation, making internal damping less likely to modify Hyperion's eccentricity. Note that in our calculations of eccentricity damping using Eq. \ref{sharma} we assumed that the energy of nutation is of the same order as total rotational energy, which may not be true for quasi-regular long-axis rotation. While longer-term alternating cycles of long-axis and chaotic rotation can be envisaged \citep{gol24b}, we have shown above that dissipation within Hyperion is in all cases greatly insufficient to arrest the secular growth of its eccentricity.

Alternatively, \citet{cuk24dda} proposed that Hyperion may be cyclically destroyed and re-accreted through a process known as ``sesquinary catastrophe'' \citep{cuk23a}. The basic idea is that Hyperion may have been captured in other, more distant (e.g, 3:2, 2:1), mean-motion resonances with Titan in the past. Due to Titan's long-term orbital expansion Hyperion's eccentricity or inclination would grow in each of the resonances. Hyperion's eccentricity would grow until ejecta originating from Hyperion would be able to re-impact Hyperion at high velocities (due to their orbits being eccentric and not aligned), leading to runaway erosion of Hyperion \citep[see ][for more explanation of sesquinary catastrophe]{cuk23a}. This would be followed by formation of a ring through collisions and re-accretion of Hyperion of a smaller orbit, to be captured into the next resonance closer to Titan. However, this mechanism cannot lead to present-day Hyperion as any eccentric orbits that would circularize close to exterior 4:3 resonance with Titan would be Titan-crossing (once out of resonance) and debris would therefore be accreted or ejected by Titan. Inclination-type resonances (necessarily of the second order) could lead to a sesquinary catastrophe of Hyperion's precursor without placing it on the collision course with Titan, but our simulations show that eccentricity-type resonances tend to be a much more common outcome of resonant encounters, even for second-order mean-motion resonances. While it is impossible to completely rule out a long history of Hyperion precursors that were sequentially captured in different resonances and destroyed through sesquinary catastrophe, numerical simulations suggest that this scenario is highly unlikely.   

\section{Limitations of using a constant rotation rate for Saturn}\label{app_b}

While the numerical integrator {\sc ssimpl} \citep{cuk18} includes the tidal evolution of Titan and precession of Saturn's spin axis due to outside torques, it does not modify the spin rate of Saturn in the process. This approximation leads to some non-conservation of angular momentum, and here we will use rough estimates of the resulting consequences to argue that our main results are not affected. 

In the process of Titan's tidal evolution the angular momentum is transferred from Saturn's spin to Titan's orbit. As Titan's semimajor axis increases by about 5\% in Fig. \ref{reverse}, we want to estimate what the resulting change in Saturn's spin rate should have been. First we estimate the ratio of angular momenta of Saturn and Titan:
\begin{equation}
{L_S \over L_T} = {\bar{C} m_S \over m_T} \Bigl({R \over a}\Bigr)^{2} {\omega_S \over n} 
\end{equation}
Where $m_S$, $R$, $\omega_S$ and $\bar{C}$ are respectively Saturn's mass, radius, spin rate and normalized polar moment of inertia, and $m_T$, $a$ and $n$ are Titan's mass, semimajor axis and mean motion. Here as in simulations shown in Fig. \ref{reverse} we use $\omega_S=5.211 \times 10^3$~rad/yr and $\bar{C}=0.2177$ \citep[cf.][]{wis22}. Using current values for other parameters, we find that $L_S=81.3 \times L_T$. 

A 5\% increase in Titan's $a$ means a 2.5\% change in $L_T$ (as $L \propto \sqrt{a}$), requiring a $3 \times 10^{-4}$ fractional decrease in Saturn's spin rate. As $J_2 \propto \omega_S^2$ \citep{md99}, Saturn's oblateness moment would decrease by a fraction of $6 \times 10^{-4}$. A decrease in spin rate would accelerate the planet's spin precession (in an inversely proportional manner), while a smaller oblateness moment would slow down axial precession \citep{war04}. As Titan's effective $J_2$ is about 6 times larger than Saturn's $J_2$ due to oblateness, precession would be both accelerated by a fraction of $3 \times 10^{-4}$ (due to a smaller $L_S$ resisting the outside torques) and decelerated by about $10^{-4}$ (due to a smaller total effective $J_2$), resulting in net $2 \times 10^{-4}$ fractional speeding up of axial precession. Given that over the same time precession accelerates by about 8\% due to Titan's orbit becoming larger (and increasing total effective $J_2$), we find the error arising from using a constant spin rate for Saturn to be negligible.  

Separately, the resulting change in the planet's $J_2$ would also have consequences on the orbital precession of the moons, notably Titan (as its secular dynamics is dominated by Saturn's oblateness). As we estimated above, this change is less than 0.1\% over the long simulation shown in Fig. \ref{reverse}. We do not expect this change to have any consequences for the scenario discussed here, especially as we propose that past instabilities were driven by mean-motion resonances rather than secular ones.


\bibliography{refs_saturn}{}

\begin{thebibliography}{}
\expandafter\ifx\csname natexlab\endcsname\relax\def\natexlab#1{#1}\fi
\providecommand{\url}[1]{\href{#1}{#1}}
\providecommand{\dodoi}[1]{doi:~\href{http://doi.org/#1}{\nolinkurl{#1}}}
\providecommand{\doeprint}[1]{\href{http://ascl.net/#1}{\nolinkurl{http://ascl.net/#1}}}
\providecommand{\doarXiv}[1]{\href{https://arxiv.org/abs/#1}{\nolinkurl{https://arxiv.org/abs/#1}}}

\bibitem[{E. {Asphaug} {et~al.}(2006){Asphaug}, {Agnor}, \& {Williams}}]{asp06}
{Asphaug}, E., {Agnor}, C.~B., \& {Williams}, Q. 2006,
  \bibinfo{title}{{Hit-and-run planetary collisions},} \nat, 439, 155,
  \dodoi{10.1038/nature04311}

\bibitem[{E. {Asphaug} \& A. {Reufer}(2013){Asphaug} \& {Reufer}}]{asp13}
{Asphaug}, E., \& {Reufer}, A. 2013, \bibinfo{title}{{Late origin of the Saturn
  system},} \icarus, 223, 544, \dodoi{10.1016/j.icarus.2012.12.009}

\bibitem[{K. {Bailli{\'e}} {et~al.}(2019){Bailli{\'e}}, {Noyelles}, {Lainey},
  {Charnoz}, \& {Tobie}}]{bai19}
{Bailli{\'e}}, K., {Noyelles}, B., {Lainey}, V., {Charnoz}, S., \& {Tobie}, G.
  2019, \bibinfo{title}{{Formation of the Cassini Division - I. Shaping the
  rings by Mimas inward migration},} \mnras, 486, 2933,
  \dodoi{10.1093/mnras/stz548}

\bibitem[{S.~W. {Bell}(2020){Bell}}]{bel20}
{Bell}, S.~W. 2020, \bibinfo{title}{{Relative Crater Scaling Between the Major
  Moons of Saturn: Implications for Planetocentric Cratering and the Surface
  Age of Titan},} Journal of Geophysical Research (Planets), 125, e06392,
  \dodoi{10.1029/2020JE006392}

\bibitem[{G.~J. {Black} {et~al.}(1995){Black}, {Nicholson}, \&
  {Thomas}}]{bla95}
{Black}, G.~J., {Nicholson}, P.~D., \& {Thomas}, P.~C. 1995,
  \bibinfo{title}{{Hyperion: Rotational dynamics.},} \icarus, 117, 149,
  \dodoi{10.1006/icar.1995.1148}

\bibitem[{D. {Brouwer} \& G.~M. {Clemence}(1961){Brouwer} \&
  {Clemence}}]{bro61}
{Brouwer}, D., \& {Clemence}, G.~M. 1961, {Methods of celestial mechanics}

\bibitem[{R.~M. {Canup} \& E. {Asphaug}(2001){Canup} \& {Asphaug}}]{can01}
{Canup}, R.~M., \& {Asphaug}, E. 2001, \bibinfo{title}{{Origin of the Moon in a
  giant impact near the end of the Earth's formation},} \nat, 412, 708,
  \dodoi{10.1038/35089010}

\bibitem[{E.~M.~A. {Chen} \& F. {Nimmo}(2016){Chen} \& {Nimmo}}]{che16}
{Chen}, E.~M.~A., \& {Nimmo}, F. 2016, \bibinfo{title}{{Tidal dissipation in
  the early lunar magma ocean and its effect on the evolution of the Earth-Moon
  system},} Icarus, 275, 132

\bibitem[{E.~M.~A. {Chen} {et~al.}(2014){Chen}, {Nimmo}, \&
  {Glatzmaier}}]{che14}
{Chen}, E.~M.~A., {Nimmo}, F., \& {Glatzmaier}, G.~A. 2014,
  \bibinfo{title}{{Tidal heating in icy satellite oceans},} Icarus, 229, 11,
  \dodoi{10.1016/j.icarus.2013.10.024}

\bibitem[{C.~F. {Chyba} {et~al.}(1989){Chyba}, {Jankowski}, \&
  {Nicholson}}]{chy89}
{Chyba}, C.~F., {Jankowski}, D.~G., \& {Nicholson}, P.~D. 1989,
  \bibinfo{title}{{Tidal evolution in the Neptune-Triton system},} Astronomy
  and Astrophysics, 219, L23

\bibitem[{A. {Crida} {et~al.}(2019){Crida}, {Charnoz}, {Hsu}, \&
  {Dones}}]{cri19}
{Crida}, A., {Charnoz}, S., {Hsu}, H.-W., \& {Dones}, L. 2019,
  \bibinfo{title}{{Are Saturn's rings actually young?},} Nature Astronomy, 3,
  967, \dodoi{10.1038/s41550-019-0876-y}

\bibitem[{M. {{\'C}uk} {et~al.}(2016){{\'C}uk}, {Dones}, \& {Nesvorn{\'
  y}}}]{cuk16}
{{\'C}uk}, M., {Dones}, L., \& {Nesvorn{\' y}}, D. 2016,
  \bibinfo{title}{{Dynamical Evidence for a Late Formation of Saturn's Moons},}
  Astrophysical Journal, 820, 97

\bibitem[{M. {{\'C}uk} {et~al.}(2013){{\'C}uk}, {Dones}, \&
  {Nesvorn{\'y}}}]{cuk13}
{{\'C}uk}, M., {Dones}, L., \& {Nesvorn{\'y}}, D. 2013,
  \bibinfo{title}{{Titan-Hyperion Resonance and the Tidal Q of Saturn},} arXiv
  e-prints, arXiv:1311.6780

\bibitem[{M. {{\'C}uk} {et~al.}(2018){{\'C}uk}, {Dones}, {Nesvorn{\'y}}, \&
  {Walsh}}]{cuk18}
{{\'C}uk}, M., {Dones}, L., {Nesvorn{\'y}}, D., \& {Walsh}, K.~J. 2018,
  \bibinfo{title}{{Secular resonance between Iapetus and the giant planets},}
  Montly Notices of the Royal Astronomical Society, 481, 5411,
  \dodoi{10.1093/mnras/sty2631}

\bibitem[{M. {{\'C}uk} \& M. {El Moutamid}(2022){{\'C}uk} \& {El
  Moutamid}}]{cuk22}
{{\'C}uk}, M., \& {El Moutamid}, M. 2022, \bibinfo{title}{{Three-body
  Resonances in the Saturnian System},} \apjl, 926, L18,
  \dodoi{10.3847/2041-8213/ac501c}

\bibitem[{M. {{\'C}uk} \& M. {El Moutamid}(2023){{\'C}uk} \& {El
  Moutamid}}]{cuk23}
{{\'C}uk}, M., \& {El Moutamid}, M. 2023, \bibinfo{title}{{A Past Episode of
  Rapid Tidal Evolution of Enceladus?},} \psj, 4, 119,
  \dodoi{10.3847/PSJ/acde80}

\bibitem[{M. {Cuk} {et~al.}(2024){Cuk}, {El Moutamid}, {Fuller}, \&
  {Lainey}}]{cuk24dda}
{Cuk}, M., {El Moutamid}, M., {Fuller}, J., \& {Lainey}, V. 2024,
  \bibinfo{title}{{Orbital Histories of Titan, Hyperion and Iapetus},} in
  AAS/Division of Dynamical Astronomy Meeting, Vol.~56, AAS/Division of
  Dynamical Astronomy Meeting, 101.01

\bibitem[{M. {{\'C}uk} {et~al.}(2023){{\'C}uk}, {Hamilton}, {Minton}, \&
  {Stewart}}]{cuk23a}
{{\'C}uk}, M., {Hamilton}, D.~P., {Minton}, D.~A., \& {Stewart}, S.~T. 2023,
  \bibinfo{title}{{Sesquinary Catastrophe for Close-in Moons with Dynamically
  Excited Orbits},} \apj, 957, 62, \dodoi{10.3847/1538-4357/acf613}

\bibitem[{J.~N. {Cuzzi} \& P.~R. {Estrada}(1998){Cuzzi} \& {Estrada}}]{cuz98}
{Cuzzi}, J.~N., \& {Estrada}, P.~R. 1998, \bibinfo{title}{{Compositional
  Evolution of Saturn's Rings Due to Meteoroid Bombardment},} \icarus, 132, 1,
  \dodoi{10.1006/icar.1997.5863}

\bibitem[{B.~G. {Downey} \& F. {Nimmo}(2025){Downey} \& {Nimmo}}]{dow25}
{Downey}, B.~G., \& {Nimmo}, F. 2025, \bibinfo{title}{{Titan's spin state as a
  constraint on tidal dissipation},} Science Advances, 11, eadl4741,
  \dodoi{10.1126/sciadv.adl4741}

\bibitem[{R.~H. {Durisen} \& P.~R. {Estrada}(2023){Durisen} \&
  {Estrada}}]{dur23}
{Durisen}, R.~H., \& {Estrada}, P.~R. 2023, \bibinfo{title}{{Large mass inflow
  rates in Saturn's rings due to ballistic transport and mass loading},}
  \icarus, 400, 115221, \dodoi{10.1016/j.icarus.2022.115221}

\bibitem[{P.~R. {Estrada} \& R.~H. {Durisen}(2023){Estrada} \&
  {Durisen}}]{est23}
{Estrada}, P.~R., \& {Durisen}, R.~H. 2023, \bibinfo{title}{{Constraints on the
  initial mass, age and lifetime of Saturn's rings from viscous evolutions that
  include pollution and transport due to micrometeoroid bombardment},} \icarus,
  400, 115296, \dodoi{10.1016/j.icarus.2022.115296}

\bibitem[{S.~N. {Ferguson} {et~al.}(2020){Ferguson}, {Rhoden}, \&
  {Kirchoff}}]{fer20}
{Ferguson}, S.~N., {Rhoden}, A.~R., \& {Kirchoff}, M.~R. 2020,
  \bibinfo{title}{{Small Impact Crater Populations on Saturn's Moon Tethys and
  Implications for Source Impactors in the System},} Journal of Geophysical
  Research (Planets), 125, e06400, \dodoi{10.1029/2020JE006400}

\bibitem[{S.~N. {Ferguson} {et~al.}(2022{\natexlab{a}}){Ferguson}, {Rhoden}, \&
  {Kirchoff}}]{fer22a}
{Ferguson}, S.~N., {Rhoden}, A.~R., \& {Kirchoff}, M.~R. 2022{\natexlab{a}},
  \bibinfo{title}{{Regional Impact Crater Mapping and Analysis on Saturn's Moon
  Dione and the Relation to Source Impactors},} Journal of Geophysical Research
  (Planets), 127, e07204, \dodoi{10.1029/2022JE007204}

\bibitem[{S.~N. {Ferguson} {et~al.}(2022{\natexlab{b}}){Ferguson}, {Rhoden},
  {Kirchoff}, \& {Salmon}}]{fer22b}
{Ferguson}, S.~N., {Rhoden}, A.~R., {Kirchoff}, M.~R., \& {Salmon}, J.~J.
  2022{\natexlab{b}}, \bibinfo{title}{{A unique Saturnian impactor population
  from elliptical craters},} Earth and Planetary Science Letters, 593, 117652,
  \dodoi{10.1016/j.epsl.2022.117652}

\bibitem[{E. {Forest} \& R.~D. {Ruth}(1990){Forest} \& {Ruth}}]{for90}
{Forest}, E., \& {Ruth}, R.~D. 1990, \bibinfo{title}{{Fourth-order symplectic
  integration},} Physica D Nonlinear Phenomena, 43, 105,
  \dodoi{10.1016/0167-2789(90)90019-L}

\bibitem[{J. {Fuller} {et~al.}(2016){Fuller}, {Luan}, \& {Quataert}}]{ful16}
{Fuller}, J., {Luan}, J., \& {Quataert}, E. 2016, \bibinfo{title}{{Resonance
  locking as the source of rapid tidal migration in the Jupiter and Saturn moon
  systems},} Montly Notices of the Royal Astronomical Society, 458, 3867,
  \dodoi{10.1093/mnras/stw609}

\bibitem[{M. {Goldberg} \& K. {Batygin}(2024{\natexlab{a}}){Goldberg} \&
  {Batygin}}]{gol24a}
{Goldberg}, M., \& {Batygin}, K. 2024{\natexlab{a}}, \bibinfo{title}{{Chaotic
  tides as a solution to the Hyperion problem},} \icarus, 413, 116014,
  \dodoi{10.1016/j.icarus.2024.116014}

\bibitem[{M. {Goldberg} \& K. {Batygin}(2024{\natexlab{b}}){Goldberg} \&
  {Batygin}}]{gol24b}
{Goldberg}, M., \& {Batygin}, K. 2024{\natexlab{b}},
  \bibinfo{title}{{Nutation-orbit resonances: The origin of the chaotic
  rotation of Hyperion and the barrel instability},} \aap, 692, A9,
  \dodoi{10.1051/0004-6361/202452443}

\bibitem[{P. {Goldreich} \& S. {Tremaine}(1982){Goldreich} \&
  {Tremaine}}]{gol82}
{Goldreich}, P., \& {Tremaine}, S. 1982, \bibinfo{title}{{The dynamics of
  planetary rings},} \araa, 20, 249,
  \dodoi{10.1146/annurev.aa.20.090182.001341}

\bibitem[{R. {Greenberg}(1973){Greenberg}}]{gre73}
{Greenberg}, R. 1973, \bibinfo{title}{{Evolution of satellite resonances by
  tidal dissipation},} \aj, 78, 338, \dodoi{10.1086/111423}

\bibitem[{D.~P. {Hamilton}(2013){Hamilton}}]{ham13}
{Hamilton}, D.~P. 2013, \bibinfo{title}{{A Late Major Merger at Saturn:
  Consequences for Titan and Iapetus},} in AAS/Division for Planetary Sciences
  Meeting Abstracts, Vol.~45, AAS/Division for Planetary Sciences Meeting
  Abstracts \#45, 302.01

\bibitem[{D.~P. {Hamilton} \& W.~R. {Ward}(2004){Hamilton} \& {Ward}}]{ham04}
{Hamilton}, D.~P., \& {Ward}, W.~R. 2004, \bibinfo{title}{{Tilting Saturn. II.
  Numerical Model},} \aj, 128, 2510, \dodoi{10.1086/424534}

\bibitem[{R. {Hyodo} \& S. {Charnoz}(2017){Hyodo} \& {Charnoz}}]{hyo17}
{Hyodo}, R., \& {Charnoz}, S. 2017, \bibinfo{title}{{Dynamical Evolution of the
  Debris Disk after a Satellite Catastrophic Disruption around Saturn},} \aj,
  154, 34, \dodoi{10.3847/1538-3881/aa74c9}

\bibitem[{R. {Hyodo} {et~al.}(2025){Hyodo}, {Genda}, \& {Madeira}}]{hyo25}
{Hyodo}, R., {Genda}, H., \& {Madeira}, G. 2025, \bibinfo{title}{{Pollution
  resistance of Saturn's ring particles during micrometeoroid impact},} Nature
  Geoscience, 18, 44, \dodoi{10.1038/s41561-024-01598-9}

\bibitem[{L. {Iess} {et~al.}(2019){Iess}, {Militzer}, {Kaspi}, {Nicholson},
  {Durante}, {Racioppa}, {Anabtawi}, {Galanti}, {Hubbard}, {Mariani},
  {Tortora}, {Wahl}, \& {Zannoni}}]{ies19}
{Iess}, L., {Militzer}, B., {Kaspi}, Y., {et~al.} 2019,
  \bibinfo{title}{{Measurement and implications of Saturn's gravity field and
  ring mass},} Science, 364, aat2965, \dodoi{10.1126/science.aat2965}

\bibitem[{R.~A. {Jacobson}(2022){Jacobson}}]{jac22}
{Jacobson}, R.~A. 2022, \bibinfo{title}{{The Orbits of the Main Saturnian
  Satellites, the Saturnian System Gravity Field, and the Orientation of
  Saturn's Pole},} \aj, 164, 199, \dodoi{10.3847/1538-3881/ac90c9}

\bibitem[{S. {Kempf} {et~al.}(2023){Kempf}, {Altobelli}, {Schmidt}, {Cuzzi},
  {Estrada}, \& {Srama}}]{kem23}
{Kempf}, S., {Altobelli}, N., {Schmidt}, J., {et~al.} 2023,
  \bibinfo{title}{{Micrometeoroid infall onto Saturn's rings constrains their
  age to no more than a few hundred million years},} Science Advances, 9,
  eadf8537, \dodoi{10.1126/sciadv.adf8537}

\bibitem[{V. {Lainey} {et~al.}(2024){Lainey}, {Rambaux}, {Tobie}, {Cooper},
  {Zhang}, {Noyelles}, \& {Bailli{\'e}}}]{lai24}
{Lainey}, V., {Rambaux}, N., {Tobie}, G., {et~al.} 2024, \bibinfo{title}{{A
  recently formed ocean inside Saturn's moon Mimas},} \nat, 626, 280,
  \dodoi{10.1038/s41586-023-06975-9}

\bibitem[{V. {Lainey} {et~al.}(2012){Lainey}, {Karatekin}, {Desmars},
  {Charnoz}, {Arlot}, {Emelyanov}, {Le Poncin-Lafitte}, {Mathis}, {Remus},
  {Tobie}, \& {Zahn}}]{lai12}
{Lainey}, V., {Karatekin}, {\"O}., {Desmars}, J., {et~al.} 2012,
  \bibinfo{title}{{Strong Tidal Dissipation in Saturn and Constraints on
  Enceladus' Thermal State from Astrometry},} Astrophysical Journal, 752, 14,
  \dodoi{10.1088/0004-637X/752/1/14}

\bibitem[{V. {Lainey} {et~al.}(2017){Lainey}, {Jacobson}, {Tajeddine},
  {Cooper}, {Murray}, {Robert}, {Tobie}, {Guillot}, {Mathis}, {Remus},
  {Desmars}, {Arlot}, {De Cuyper}, {Dehant}, {Pascu}, {Thuillot}, {Le
  Poncin-Lafitte}, \& {Zahn}}]{lai17}
{Lainey}, V., {Jacobson}, R.~A., {Tajeddine}, R., {et~al.} 2017,
  \bibinfo{title}{{New constraints on Saturn's interior from Cassini
  astrometric data},} Icarus, 281, 286, \dodoi{10.1016/j.icarus.2016.07.014}

\bibitem[{V. {Lainey} {et~al.}(2020){Lainey}, {Casajus}, {Fuller}, {Zannoni},
  {Tortora}, {Cooper}, {Murray}, {Modenini}, {Park}, {Robert}, \&
  {Zhang}}]{lai20}
{Lainey}, V., {Casajus}, L.~G., {Fuller}, J., {et~al.} 2020,
  \bibinfo{title}{{Resonance locking in giant planets indicated by the rapid
  orbital expansion of Titan},} Nature Astronomy, 4, 1053,
  \dodoi{10.1038/s41550-020-1120-5}

\bibitem[{J.~J. {Lissauer} {et~al.}(1985){Lissauer}, {Goldreich}, \&
  {Tremaine}}]{lis85b}
{Lissauer}, J.~J., {Goldreich}, P., \& {Tremaine}, S. 1985,
  \bibinfo{title}{{Evolution of the Janus-Epimetheus coorbital resonance due to
  torques from Saturn's ring},} \icarus, 64, 425,
  \dodoi{10.1016/0019-1035(85)90066-1}

\bibitem[{K.~E. {Mandt} {et~al.}(2012){Mandt}, {Waite}, {Teolis}, {Magee},
  {Bell}, {Westlake}, {Nixon}, {Mousis}, \& {Lunine}}]{man12}
{Mandt}, K.~E., {Waite}, J.~H., {Teolis}, B., {et~al.} 2012,
  \bibinfo{title}{{The $^{12}$C/$^{13}$C Ratio on Titan from Cassini INMS
  Measurements and Implications for the Evolution of Methane},} \apj, 749, 160,
  \dodoi{10.1088/0004-637X/749/2/160}

\bibitem[{C.~R. {Mankovich} {et~al.}(2023){Mankovich}, {Dewberry}, \&
  {Fuller}}]{man23}
{Mankovich}, C.~R., {Dewberry}, J.~W., \& {Fuller}, J. 2023,
  \bibinfo{title}{{Saturn's Seismic Rotation Revisited},} \psj, 4, 59,
  \dodoi{10.3847/PSJ/acc253}

\bibitem[{S. {Millholland} \& G. {Laughlin}(2019){Millholland} \&
  {Laughlin}}]{mil19}
{Millholland}, S., \& {Laughlin}, G. 2019, \bibinfo{title}{{Obliquity-driven
  sculpting of exoplanetary systems},} Nature Astronomy, 3, 424,
  \dodoi{10.1038/s41550-019-0701-7}

\bibitem[{C.~D. Murray \& S.~F. Dermott(1999)Murray \& Dermott}]{md99}
Murray, C.~D., \& Dermott, S.~F. 1999, Solar System Dynamics (Cambridge
  University Press)

\bibitem[{C.~D. {Neish} \& R.~D. {Lorenz}(2012){Neish} \& {Lorenz}}]{nei12}
{Neish}, C.~D., \& {Lorenz}, R.~D. 2012,
  \bibinfo{title}{{Titan{\textquoteright}s global crater population: A new
  assessment},} \planss, 60, 26, \dodoi{10.1016/j.pss.2011.02.016}

\bibitem[{F. {Petricca} {et~al.}(2025){Petricca}, {Vance}, {Parisi}, ,
  {Buccino}, {Cascioli}, {CAstillo-Rogez}, {Downey}, {Nimmo}, {Tobie},
  {Journaux}, {Magnanini}, {Jones}, {Panning}, {Bagheri}, {Genova}, \&
  {Lunine}}]{pet25}
{Petricca}, F., {Vance}, S.~D., {Parisi}, M., {et~al.} 2025,
  \bibinfo{title}{{Titan’s strong tidal dissipation precludes a subsurface
  ocean},} \nat, 648, 556

\bibitem[{W. {Polycarpe} {et~al.}(2018){Polycarpe}, {Saillenfest}, {Lainey},
  {Vienne}, {Noyelles}, \& {Rambaux}}]{pol18}
{Polycarpe}, W., {Saillenfest}, M., {Lainey}, V., {et~al.} 2018,
  \bibinfo{title}{{Strong tidal energy dissipation in Saturn at Titan's
  frequency as an explanation for Iapetus orbit},} \aap, 619, A133,
  \dodoi{10.1051/0004-6361/201833930}

\bibitem[{A.~R. {Rhoden}(2023){Rhoden}}]{rho23}
{Rhoden}, A.~R. 2023, \bibinfo{title}{{Mimas: Frozen Fragment, Ring Relic, or
  Emerging Ocean World?},} Annual Review of Earth and Planetary Sciences, 51,
  367, \dodoi{10.1146/annurev-earth-031621-061221}

\bibitem[{A.~R. {Rhoden} \& M.~E. {Walker}(2022){Rhoden} \& {Walker}}]{rho22}
{Rhoden}, A.~R., \& {Walker}, M.~E. 2022, \bibinfo{title}{{The case for an
  ocean-bearing Mimas from tidal heating analysis},} \icarus, 376, 114872,
  \dodoi{10.1016/j.icarus.2021.114872}

\bibitem[{A.~R. {Rhoden} {et~al.}(2024){Rhoden}, {Walker}, {Rudolph}, {Bland},
  \& {Manga}}]{rho24a}
{Rhoden}, A.~R., {Walker}, M.~E., {Rudolph}, M.~L., {Bland}, M.~T., \& {Manga},
  M. 2024, \bibinfo{title}{{The evolution of a young ocean within Mimas},}
  Earth and Planetary Science Letters, 635, 118689,
  \dodoi{10.1016/j.epsl.2024.118689}

\bibitem[{S.~J. {Robbins} {et~al.}(2024){Robbins}, {Bierhaus}, \&
  {Dones}}]{rob24}
{Robbins}, S.~J., {Bierhaus}, E.~B., \& {Dones}, L. 2024,
  \bibinfo{title}{{Crater Populations of the Saturnian Satellites Mimas, Rhea,
  and Iapetus},} Journal of Geophysical Research (Planets), 129, e2023JE007941,
  \dodoi{10.1029/2023JE007941}

\bibitem[{S.~J. {Robbins} \& K.~N. {Singer}(2021){Robbins} \& {Singer}}]{rob21}
{Robbins}, S.~J., \& {Singer}, K.~N. 2021, \bibinfo{title}{{Pluto and Charon
  Impact Crater Populations: Reconciling Different Results},} \psj, 2, 192,
  \dodoi{10.3847/PSJ/ac0e94}

\bibitem[{C. {Sagan} \& S.~F. {Dermott}(1982){Sagan} \& {Dermott}}]{sag82}
{Sagan}, C., \& {Dermott}, S.~F. 1982, \bibinfo{title}{{The tide in the seas of
  Titan},} \nat, 300, 731, \dodoi{10.1038/300731a0}

\bibitem[{M. {Saillenfest} {et~al.}(2021{\natexlab{a}}){Saillenfest}, {Lari},
  \& {Bou{\'e}}}]{sai21b}
{Saillenfest}, M., {Lari}, G., \& {Bou{\'e}}, G. 2021{\natexlab{a}},
  \bibinfo{title}{{The large obliquity of Saturn explained by the fast
  migration of Titan},} Nature Astronomy, 5, 345,
  \dodoi{10.1038/s41550-020-01284-x}

\bibitem[{M. {Saillenfest} {et~al.}(2021{\natexlab{b}}){Saillenfest}, {Lari},
  {Bou{\'e}}, \& {Courtot}}]{sai21a}
{Saillenfest}, M., {Lari}, G., {Bou{\'e}}, G., \& {Courtot}, A.
  2021{\natexlab{b}}, \bibinfo{title}{{The past and future obliquity of Saturn
  as Titan migrates},} \aap, 647, A92, \dodoi{10.1051/0004-6361/202039891}

\bibitem[{I. {Sharma} {et~al.}(2005){Sharma}, {Burns}, \& {Hui}}]{sha05}
{Sharma}, I., {Burns}, J.~A., \& {Hui}, C.~Y. 2005, \bibinfo{title}{{Nutational
  damping times in solids of revolution},} \mnras, 359, 79,
  \dodoi{10.1111/j.1365-2966.2005.08864.x}

\bibitem[{K.~N. {Singer} {et~al.}(2019){Singer}, {McKinnon}, {Gladman},
  {Greenstreet}, {Bierhaus}, {Stern}, {Parker}, {Robbins}, {Schenk}, {Grundy},
  {Bray}, {Beyer}, {Binzel}, {Weaver}, {Young}, {Spencer}, {Kavelaars},
  {Moore}, {Zangari}, {Olkin}, {Lauer}, {Lisse}, {Ennico}, {New Horizons
  Geology}, Team, {New Horizons Surface Composition Science Theme Team}, \&
  {New Horizons Ralph and LORRI Teams}}]{sin19}
{Singer}, K.~N., {McKinnon}, W.~B., {Gladman}, B., {et~al.} 2019,
  \bibinfo{title}{{Impact craters on Pluto and Charon indicate a deficit of
  small Kuiper belt objects},} Science, 363, 955,
  \dodoi{10.1126/science.aap8628}

\bibitem[{R. {Tajeddine} {et~al.}(2017){Tajeddine}, {Nicholson}, {Longaretti},
  {El Moutamid}, \& {Burns}}]{taj17}
{Tajeddine}, R., {Nicholson}, P.~D., {Longaretti}, P.-Y., {El Moutamid}, M., \&
  {Burns}, J.~A. 2017, \bibinfo{title}{{What Confines the Rings of Saturn?},}
  \apjs, 232, 28, \dodoi{10.3847/1538-4365/aa8c09}

\bibitem[{L.~F.~A. {Teodoro} {et~al.}(2023){Teodoro}, {Kegerreis}, {Estrada},
  {{\'C}uk}, {Eke}, {Cuzzi}, {Massey}, \& {Sandnes}}]{teo23}
{Teodoro}, L.~F.~A., {Kegerreis}, J.~A., {Estrada}, P.~R., {et~al.} 2023,
  \bibinfo{title}{{A Recent Impact Origin of Saturn's Rings and Mid-sized
  Moons},} \apj, 955, 137, \dodoi{10.3847/1538-4357/acf4ed}

\bibitem[{P.~C. {Thomas} {et~al.}(2007){Thomas}, {Armstrong}, {Asmar}, {Burns},
  {Denk}, {Giese}, {Helfenstein}, {Iess}, {Johnson}, {McEwen}, {Nicolaisen},
  {Porco}, {Rappaport}, {Richardson}, {Somenzi}, {Tortora}, {Turtle}, \&
  {Veverka}}]{tho07}
{Thomas}, P.~C., {Armstrong}, J.~W., {Asmar}, S.~W., {et~al.} 2007,
  \bibinfo{title}{{Hyperion's sponge-like appearance},} \nat, 448, 50,
  \dodoi{10.1038/nature05779}

\bibitem[{P. {Tortora} {et~al.}(2016){Tortora}, {Zannoni}, {Hemingway},
  {Nimmo}, {Jacobson}, {Iess}, \& {Parisi}}]{tor16}
{Tortora}, P., {Zannoni}, M., {Hemingway}, D., {et~al.} 2016,
  \bibinfo{title}{{Rhea gravity field and interior modeling from Cassini data
  analysis},} \icarus, 264, 264, \dodoi{10.1016/j.icarus.2015.09.022}

\bibitem[{J. Touma \& J. Wisdom(1998)Touma \& Wisdom}]{tou98}
Touma, J., \& Wisdom, J. 1998, \bibinfo{title}{{Resonances in the early
  evolution of the earth-moon system},} Astronomical Journal, 115, 1653

\bibitem[{R.~H. {Tyler}(2008){Tyler}}]{tyl08}
{Tyler}, R.~H. 2008, \bibinfo{title}{{Strong ocean tidal flow and heating on
  moons of the outer planets},} Nature, 456, 770, \dodoi{10.1038/nature07571}

\bibitem[{W.~R. {Ward} \& D.~P. {Hamilton}(2004){Ward} \& {Hamilton}}]{war04}
{Ward}, W.~R., \& {Hamilton}, D.~P. 2004, \bibinfo{title}{{Tilting Saturn. I.
  Analytic Model},} \aj, 128, 2501, \dodoi{10.1086/424533}

\bibitem[{J. {Wisdom} {et~al.}(2022){Wisdom}, {Dbouk}, {Militzer}, {Hubbard},
  {Nimmo}, {Downey}, \& {French}}]{wis22}
{Wisdom}, J., {Dbouk}, R., {Militzer}, B., {et~al.} 2022, \bibinfo{title}{{Loss
  of a satellite could explain Saturn{\textquoteright}s obliquity and young
  rings},} Science, 377, 1285

\bibitem[{J. {Wisdom} {et~al.}(1984){Wisdom}, {Peale}, \& {Mignard}}]{wis84}
{Wisdom}, J., {Peale}, S.~J., \& {Mignard}, F. 1984, \bibinfo{title}{{The
  chaotic rotation of Hyperion},} \icarus, 58, 137,
  \dodoi{10.1016/0019-1035(84)90032-0}

\bibitem[{K. {Zahnle} {et~al.}(2003){Zahnle}, {Schenk}, {Levison}, \&
  {Dones}}]{zah03}
{Zahnle}, K., {Schenk}, P., {Levison}, H., \& {Dones}, L. 2003,
  \bibinfo{title}{{Cratering rates in the outer Solar System},} \icarus, 163,
  263, \dodoi{10.1016/S0019-1035(03)00048-4}

\end{thebibliography}
\bibliographystyle{aasjournalv7}



\end{document}